\documentclass[]{mn2e}
\usepackage{graphicx}
\usepackage{epsfig}
\usepackage{amsmath}
\newcommand       \Angstrom     {\,{\rm \AA}}

\newcommand       \cm           {\,{\rm cm}}

\newcommand       \km           {\,{\rm km}}
\newcommand       \erg          {\,{\rm erg}}

\newcommand       \s            {\,{\rm s}}
\newcommand       \Hz         {\,{\rm Hz}}

\newcommand     \gtsim  {\lower.5ex\hbox{$\buildrel > \over \sim$}}
\newcommand     \ltsim  {\lower.5ex\hbox{$\buildrel < \over \sim$}}
\newcommand     \simgt  {\lower.5ex\hbox{$\buildrel > \over \sim$}}
\newcommand     \simlt  {\lower.5ex\hbox{$\buildrel < \over \sim$}}

\newcommand       \mum          {\,{\rm \mu m}}

\newcommand       \Teff         {T_{\rm eff}}

\newcommand       \simali       {\sim\,}

\newcommand       \Fobs          {F_\lambda^{\rm obs}}
\newcommand       \Fintrinsic   {F_\lambda^{\rm int}}
\newcommand       \Frk             {F_\nu^{\rm RK}}
\title{
The 2175$\Angstrom$ Interstellar Extinction Bump:
Is the Wavelength Variable?
}
\author[Wang, Yang \& Li]
            {Qian~Wang$^{1,2}$,
              X.J.~Yang$^{1,2}$\thanks{xjyang@xtu.edu.cn},
             and Aigen Li$^{2}$\thanks{lia@missouri.edu}\\
 $^1$Hunan Key Laboratory for Stellar
              and Interstellar Physics
              and School of Physics and Optoelectronics,\\
              Xiangtan University, Hunan 411105, China\\
  $^2$Department of Physics and Astronomy,
                  University of Missouri,
                  Columbia, MO 65211, USA\\
                  }

\begin{document}

\date{}
\pagerange{\pageref{firstpage}--\pageref{lastpage}} \pubyear{2023}

\maketitle

\label{firstpage}
\begin{abstract}
The most striking characteristics of
the mysterious 2175$\Angstrom$
extinction bump, the strongest spectroscopic
absorption feature seen on the interstellar
extinction curve, are the invariant central
wavelength and variable bandwidth:
its peak position at 2175$\Angstrom$
is remarkably constant while its bandwidth
varies from one line of sight to another.
However, recent studies of the lines of sight
toward a number of Herbig Ae/Be stars have
revealed that the extinction bump exhibits
substantial shifts from the canonical wavelength
of 2175$\Angstrom$.
In this work we revisit these lines of sight
and take a physical approach to determine
the ultraviolet (UV) extinction curve for each
line of sight. It is found that the wavelengths
of the derived UV extinction bumps are around
2200$\Angstrom$ and the scatters are considerably
smaller than that of the previous study based on
the same set of Herbig Ae/Be stars,
consistent with the conventional wisdom.
Nevertheless, the scatters are still appreciably
larger than that associated with the nominal
bump position of 2175$\Angstrom$.
This is discussed in the context that
Herbig Ae/Be stars are not well-suited
for interstellar extinction studies.
\end{abstract}

\section{Introduction}\label{sec:intro}
The 2175$\Angstrom$ extinction bump
was first detected nearly six decades ago
by Stecher (1965).
It shows up as the most prominent spectral feature
on the interstellar extinction curve.
It is widely seen in the Milky Way and nearby galaxies,
including the Large Magellanic Cloud,
several regions in the Small Magellanic Cloud,
and M31 (see Whittet 2022).
Very recently, it was detected by
the {\it James Webb Space Telescope} (JWST)
in a distant galaxy at redshift $z\approx6.71$
(Witstok et al.\ 2023).

Despite nearly 60 years' extensive observational,
theoretical and experimental studies, the exact
carrier of the 2175$\Angstrom$ extinction bump
remains unidentified, although various candidate
materials have been proposed, including small
graphitic grains (Stecher \& Donn 1965, Draine 1988),
polycyclic aromatic hydrocarbon
(PAH; Joblin et al.\ 1992, Li \& Draine 2001,
Steglich et al.\ 2013), and carbon buckyonions
composed of spherical concentric fullerene shells
(Chhowalla et al.\ 2003,  Iglesias-Groth et al.\ 2003,
Ruiz et al.\ 2005, Li et al.\ 2008).
More recently, Ma et al.\ (2020) found that
a new carbon allotrope known as T-carbon,
formed by substituting each atom in diamond
with a carbon tetrahedron (Sheng et al.\ 2011),
exhibits a prominent peak around 2175$\Angstrom$
in its electronic absorption spectrum.

The difficulty in identifying the carrier of
the 2175$\Angstrom$ extinction bump
is mainly related to its striking characteristics:
while its strength and width vary with environment,
its peak position is quite invariant:
the central wavelength of this feature varies by only
$\pm$0.46\% (2$\sigma$) around 2175$\Angstrom$
(4.6$\mum^{-1}$), while its full width half maximum
(FWHM) varies by $\pm$12\% (2$\sigma$)
around 469$\Angstrom$ ($\approx$1$\mum^{-1}$).
The bump width appears to show a strong correlation
with the environment. In general, broad bumps are
often seen in dense, quiescent environments,
while narrower bumps are mainly seen in diffuse regions
and regions of recent early-type star formation
(Fitzpatrick \& Massa 1986; hereafter FM86).
There is no substance that can explain
the observational fact that the central wavelength
of the 2175$\Angstrom$ extinction bump basically
does not change with the environment.
For example, although increasing the size of
graphite grains can broaden the bump,
its peak shifts to longer wavelength
which is inconsistent with the observational
characteristic of a stable peak wavelength
(Draine \& Malhotra 1993, Mathis 1994).

Recently, Blasberger et al.\ (2017) analyzed
the UV spectra of 26 Herbig Ae/Be (HAeBe) stars
obtained with the {\it International Ultraviolet Explorer}
(IUE) and determined the UV extinction curves
for the lines of sight toward these stars.
Contrary to the conventional wisdom,
the extinction bumps derived by Blasberger et al.\ (2017)
do not always peak at 2175$\Angstrom$.
Instead, they exhibit significant shifts
of the central wavelength of the extinction bump.
If confirmed, this would be an important breakthrough
in the interstellar extinction studies,
particularly, this will provide valuable
insight into the carriers of the 2175$\Angstrom$ bump.
This motivates us to revisit all these 26 sight lines.
In \S\ref{sec:methods} we briefly describe the sample
and method. The results are presented and discussed
in \S\ref{sec:results}. Our major conclusion is summarized
in \S\ref{sec:summary}.

\section{The Sample and Method}\label{sec:methods}
We consider the sample of Blasberger et al.\ (2017),
which consists of 26 pre-main-sequence stars
of spectral type A2 and earlier.
All have significant UV absorption
(equivalent width, $\mathrm{EW}\simgt100\Angstrom$)
around 2175$\Angstrom$.
The sample and the stellar parameters are listed
in Table~\ref{tab:stellarpara}.

Following Blasberger et al.\ (2017),
we derive the UV extinction curve
for each line of sight by comparing $\Fobs$,
the stellar spectrum observed by the IUE
which suffers dust extinction,
with $\Fintrinsic$, the {\it intrinsic} stellar spectrum
which is free of dust extinction.
The IUE spectra were extracted from
the {\it Mikulski Archive for Space Telescopes},\footnote{
    {\sf https://archive.stsci.edu/iue/}
  }
by combining data obtained with three cameras:
the short-wavelength camera (SWP)
which spans the wavelength range
of 1200--1970$\Angstrom$,
and the two long-wavelength cameras (LWP and LWR)
which span the wavelength range
of 1970--3200$\Angstrom$.

We approximate the intrinsic stellar spectrum
by the stellar model atmospheric spectrum
of Castelli \& Kurucz (2004).\footnote{%
  {\sf http://kurucz.harvard.edu/grids/gridxxxodfnew}
  }
Let $A_{\lambda}$ be the extinction
at wavelength $\lambda$, and
$A_V$ be the visual extinction.
The observed IUE spectrum
and the intrinsic stellar spectrum
are related through dust extinction
\begin{equation}\label{eq:fit}
\Fobs = \Fintrinsic
\exp \left(-\frac{A_{\lambda}}{A_{V}} \frac{A_{V}}{1.086}\right) ~.
\end{equation}
Let $\Frk$ be the Kurucz model
atmospheric flux at stellar surface
(in unit of $\erg\s^{-1}\cm^{-2}\Hz^{-1}$).
At an Earth-stellar distance of $d$,
the intrinsic, extinction-free flux
(in unit of $\erg\s^{-1}\cm^{-3}$)
would be
\begin{equation}
\Fintrinsic = \Frk \times\frac{c}{\lambda^2}
\times \left(\frac{R_\star}{d}\right)^2 ~,
\end{equation}
where $c$ is the speed of light.

%

Following Fitzpatrick \& Massa (1988; hereafter FM88)
and Cardelli et al.\ (1989),
we represent the wavelength-dependence
of extinction by an analytical formula
consisting of three parts,
\begin{equation}
\frac{A_{\lambda}}{A_{V}} = c_{1} + c_{2} x
+ c_{3} D\left(x; \gamma, x_{0}\right) + c_{4} F(x) ~,
\end{equation}
where $x\equiv \lambda^{-1}$ is the inverse wavelength;
$c_{1} + c_{2} x$ is the linear background;
$D\left(x;\gamma, x_{0}\right)$, a Drude function
of width $\gamma$ (in unit of $\mum^{-1}$)
peaking at $x_0$ (also in unit of $\mum^{-1}$),
characterizes the 2175$\Angstrom$ extinction
bump and is defined as
\begin{equation}
D\left(\mathrm{x} ; \gamma, x_{0}\right)
=\frac{x^{2}}{\left(x^{2}-x_{0}^{2}\right)^{2}+x^{2} \gamma^{2}} ~,
\end{equation}
and $F(x)$ is the far-UV (FUV) nolinear rise
at $x>5.9\mum^{-1}$ as described by
\begin{equation}
  F(x)=
  \begin{cases}
    0.5392(x-5.9)^{2}+0.05644(x-5.9)^{3}
    & x \geq 5.9\,\mu\mathrm{m}^{-1}  ~,\\
    0 & \mathrm{x}<5.9\,\mu\mathrm{m}^{-1} ~.
  \end{cases}
\end{equation}
%


The Castelli \& Kurucz (2004) stellar atmospheric
model is characterized by four parameters:
effective temperature $\Teff$,
gravity log\,$g$,
metal abundance ${\rm \left[M/H\right]}$,
and microturbulence $\xi$.
The Castelli \& Kurucz (2004) stellar
model spectral library spans a range
in $\Teff$ from 35,00\,K to 50,000\,K,
${\rm \left[M/H\right]}
= -1.5, -1.0, -0.5, 0.0$, and $0.5$
(relative to solar abundance),
log\,$g$ from 0.0 dex to 5.0 dex,
and $\xi$\,=\,2.0$\km\s^{-1}$.
For a source of given spectral type
(therefore $\Teff$), to select the most appropriate
stellar model atmospheric spectrum,
we try all the possible metallicities and gravities
and select those that best fit the measured spectrum.

For each object, we make use of
the Levenberg–Marquardt algorithm to fit
the observed spectrum
(see eq.\,\ref{eq:fit})
by minimizing ${\chi }^{2}$.
The best-fit parameters
$A_V$, $\gamma$, $x_0$, $c_1$, $c_2$,
$c_3$, and $c_4$ are listed in Table~\ref{tab:modpara}.
During the fitting process, we find that
the bump width $\gamma$ and peak position
$x_0$ are relatively independent of the other
parameters and can be accurately determined.
Therefore, we first fix $\gamma$ and $x_0$
and fit the observed IUE spectrum to search
for best-fitting $c_1$, $c_2$ and $c_4$.
We then fix $\gamma$, $x_0$, $c_1$, $c_2$
and $c_4$ and re-fit the observed spectrum
to determine $c_3$ and $A_V$.
Finally, we take these values as initial
``guesses'' and fit the observed spectrum
again so that we derive a full set of
model parameters (i.e., $A_V$, $c_1$, $c_2$,
$\gamma$, $x_0$, $c_3$, and $c_4$).
We repeat this last step three times
so that we achieve the minimum $\chi^2$.

\section{Results and Discussion}\label{sec:results}
\begin{figure*}
\vspace{-5mm}
\begin{center}
\includegraphics[width=16cm,angle=0]{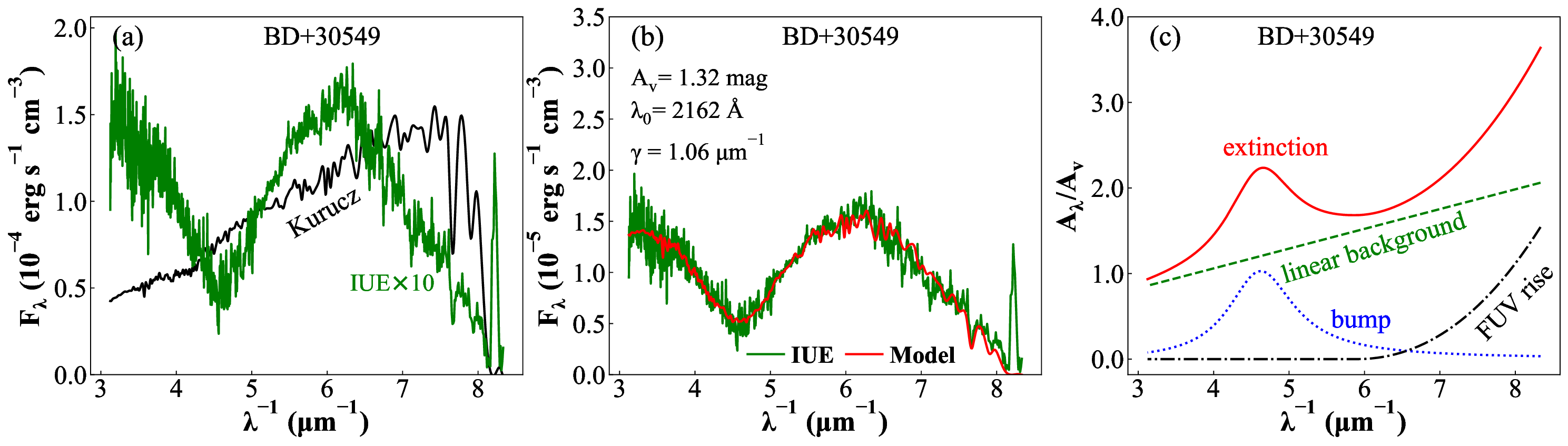}
\vspace{-2mm}
\includegraphics[width=16cm,angle=0]{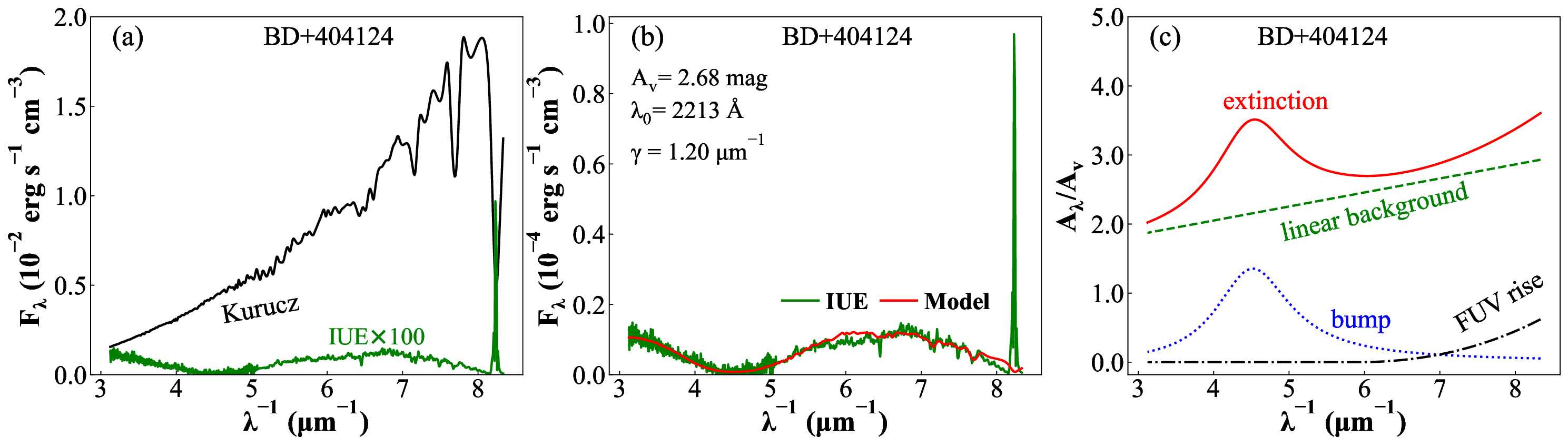}
\vspace{-1mm}
\includegraphics[width=16cm,angle=0]{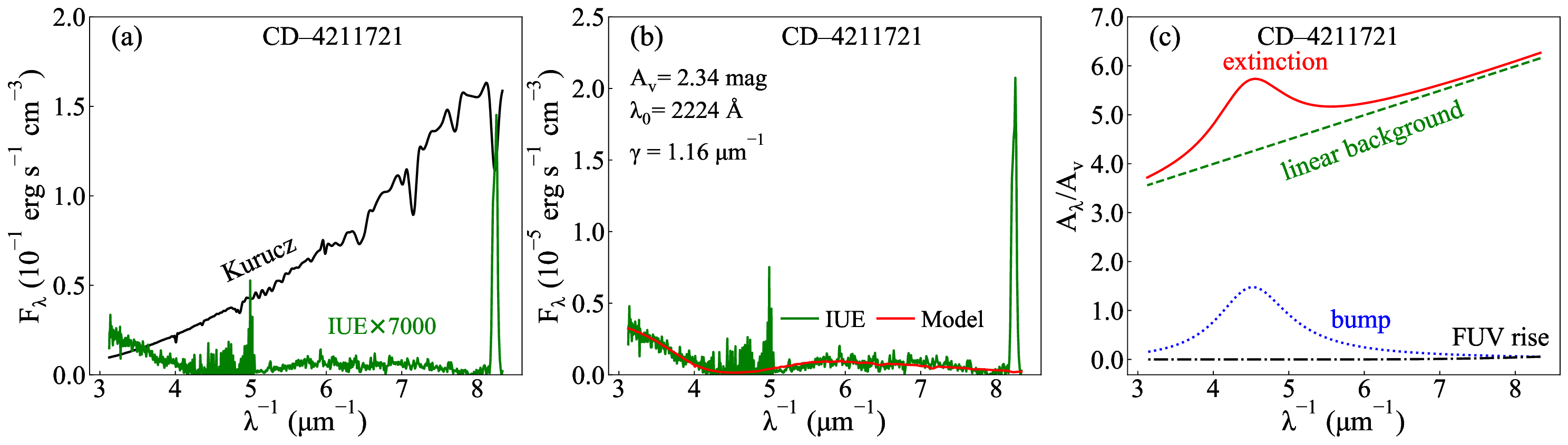}
\vspace{-1mm}
\includegraphics[width=16cm,angle=0]{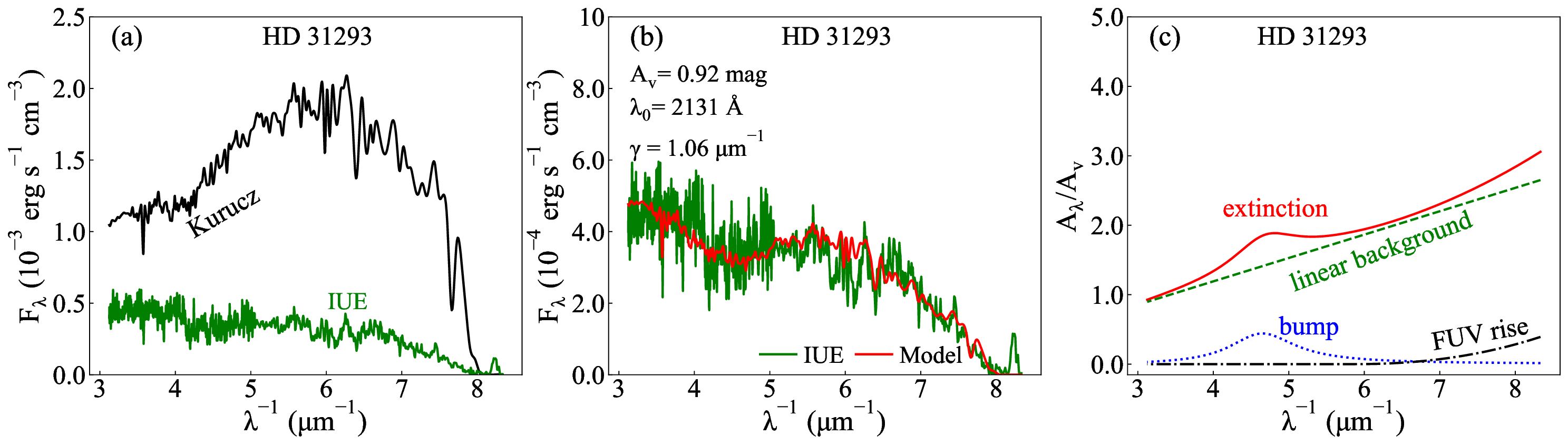}
\vspace{-1mm}
\includegraphics[width=16cm,angle=0]{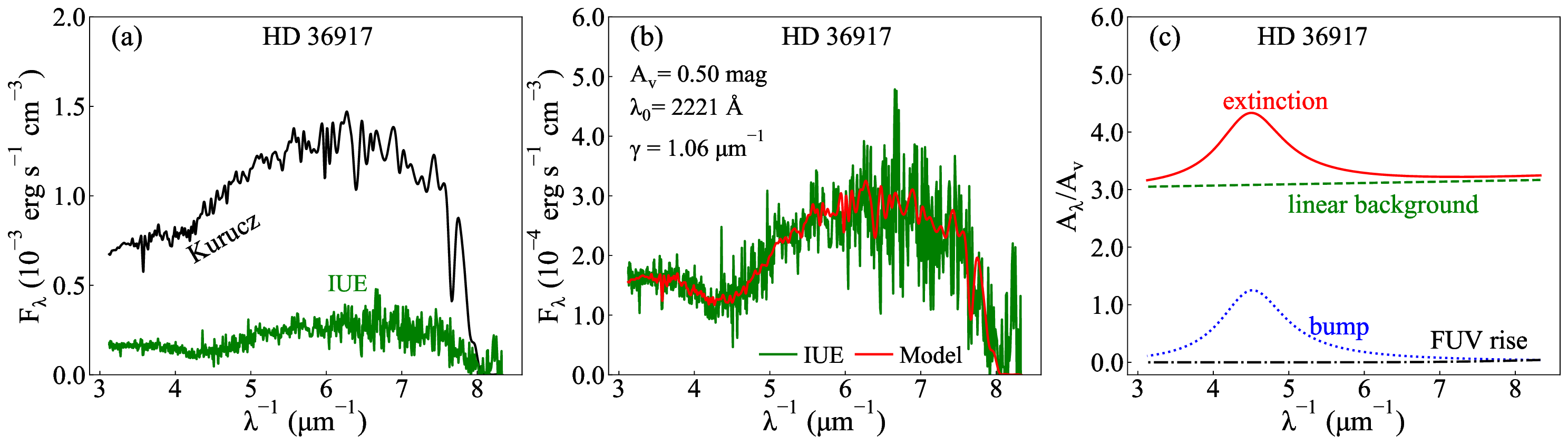}
\end{center}
\vspace{-4mm}
\caption{\label{fig:modfit1} 
  Observed, extinction-suffering IUE spectra ($\Fobs$),
  ``intrinsic'', extinction-free spectra ($\Fintrinsic$),
  and UV extinction curves of the interstellar lines of sight
  toward BD+30549, BD+404124, CD-4211721, HD~31293, and HD~36917.
  The left columns compare $\Fobs$ (green solid lines)
  with $\Fintrinsic$ (black solid lines).
  The middle columns highlight the best fits
  (red solid lines) to the observed spectra (green solid lines).
  The right columns show the interstellar extinction curves
  expressed as $A_{\lambda}$/$A_V$.
  The red line is the FM parametrization
  at $\lambda^{-1} > 3.3\mum^{-1}$,
  which is the sum of a linear ``background'' (green line),
  a Drude bump of width $\gamma$ peaking at
  $x_0\equiv\lambda_0^{-1}$ (blue line),
  and a nonlinear FUV rise (black dash-dotted line)
  at $\lambda^{-1}>5.9\mum^{-1}$.
	 }
\vspace{-3mm}
\end{figure*}

\begin{figure*}
\vspace{-3mm}
\begin{center}
\includegraphics[width=16cm,angle=0]{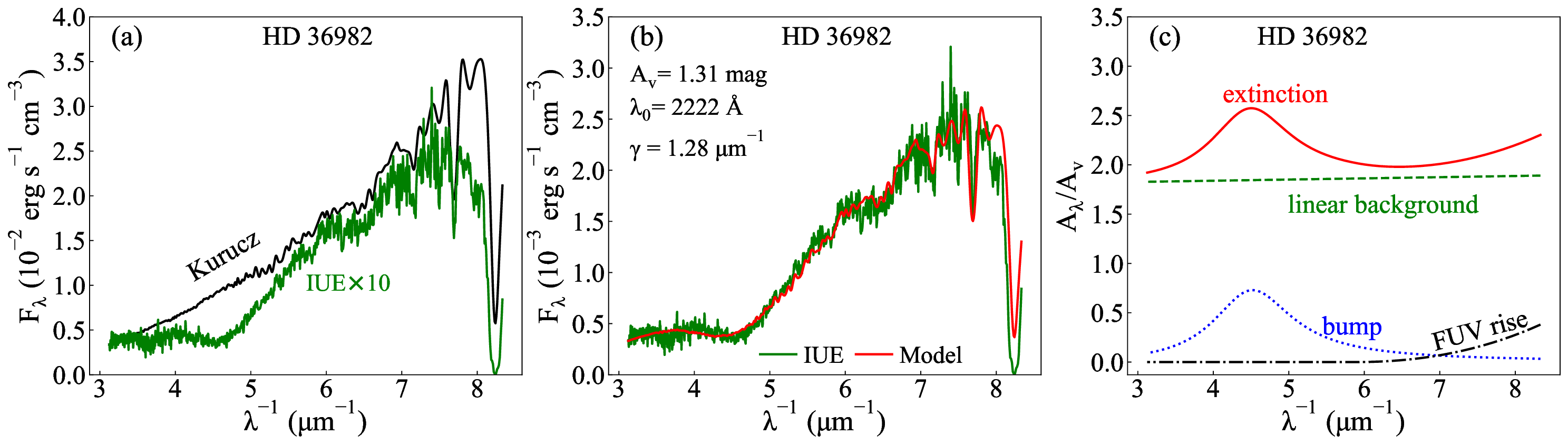}
\includegraphics[width=16cm,angle=0]{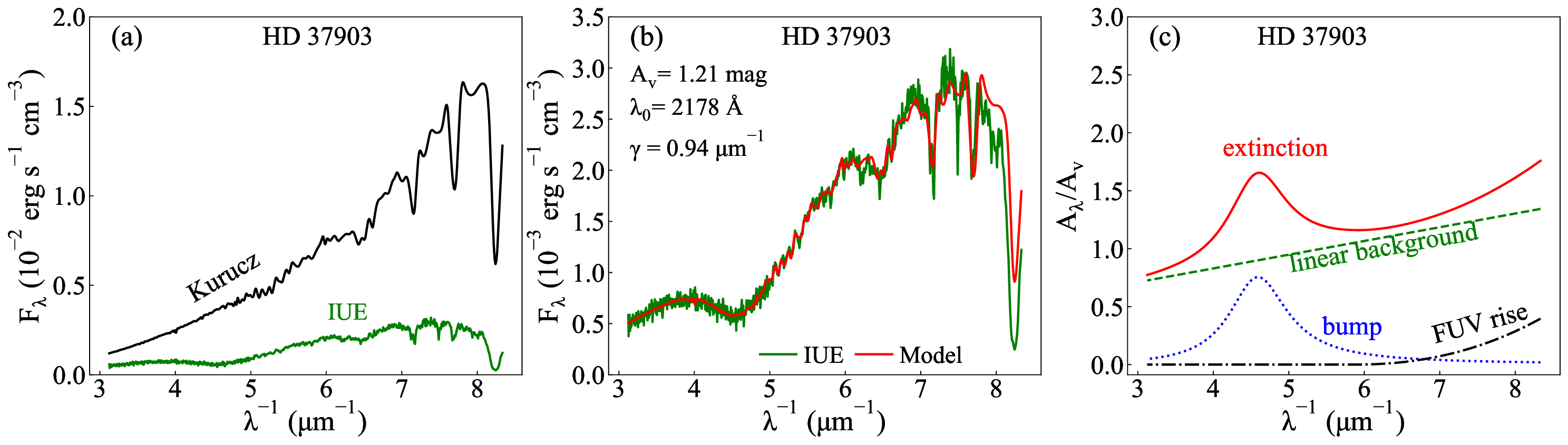}
\includegraphics[width=16cm,angle=0]{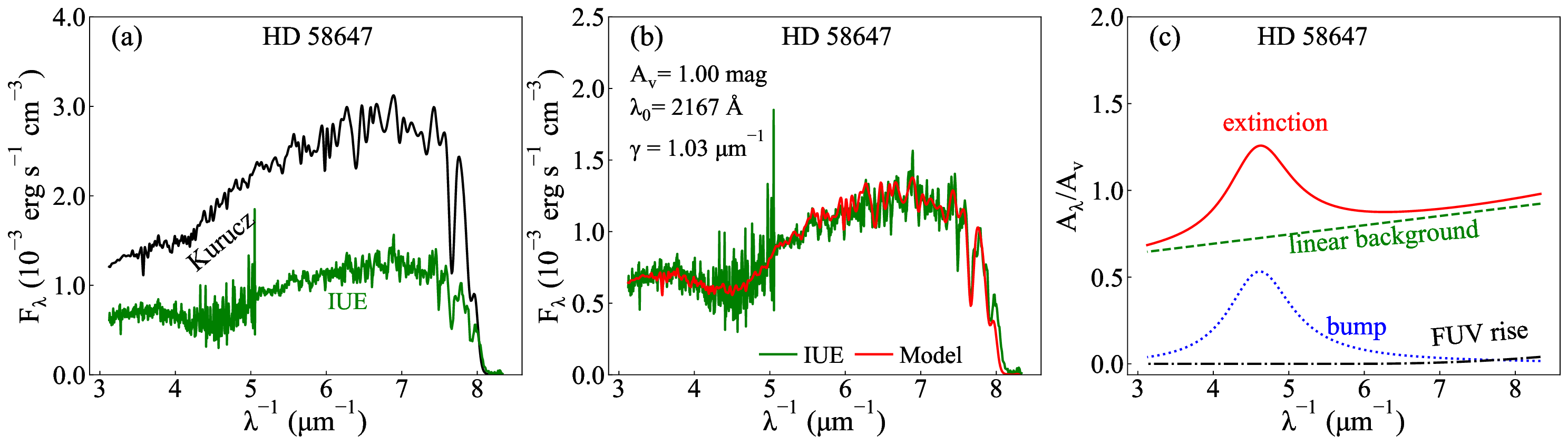}
\includegraphics[width=16cm,angle=0]{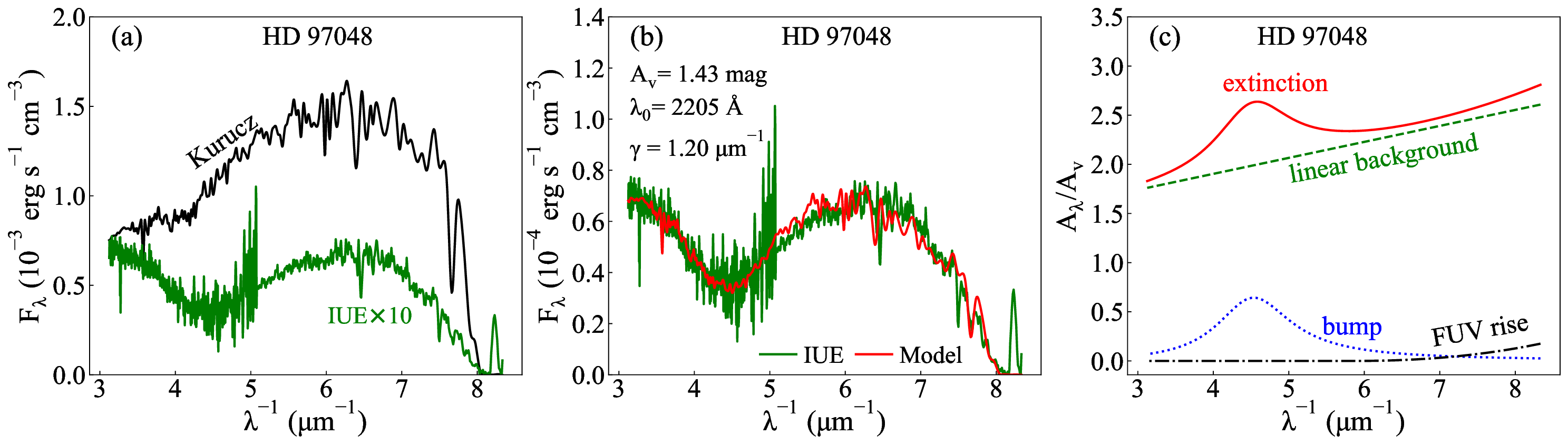}
\includegraphics[width=16cm,angle=0]{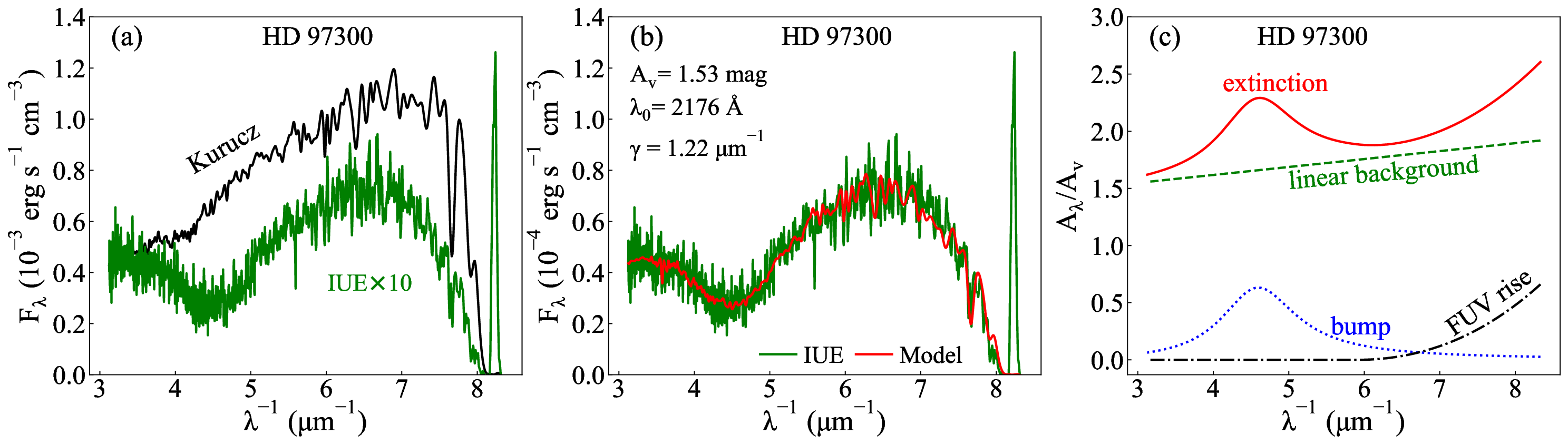}
\end{center}
\vspace{-5mm}
\caption{\label{fig:modfit2} 
  Same as Figure~\ref{fig:modfit1}
  but for HD~36982, HD~37903, HD~58647, HD~97048, and HD~97300
	 }
\vspace{-3mm}
\end{figure*}

\begin{figure*}
\vspace{-3mm}
\begin{center}
\includegraphics[width=16cm,angle=0]{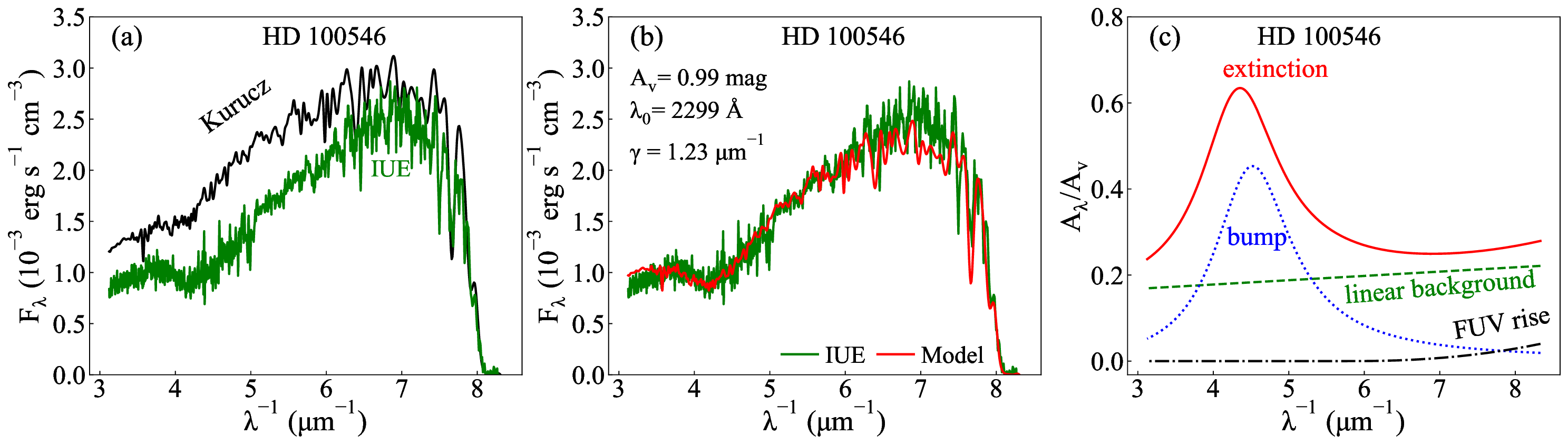}
\includegraphics[width=16cm,angle=0]{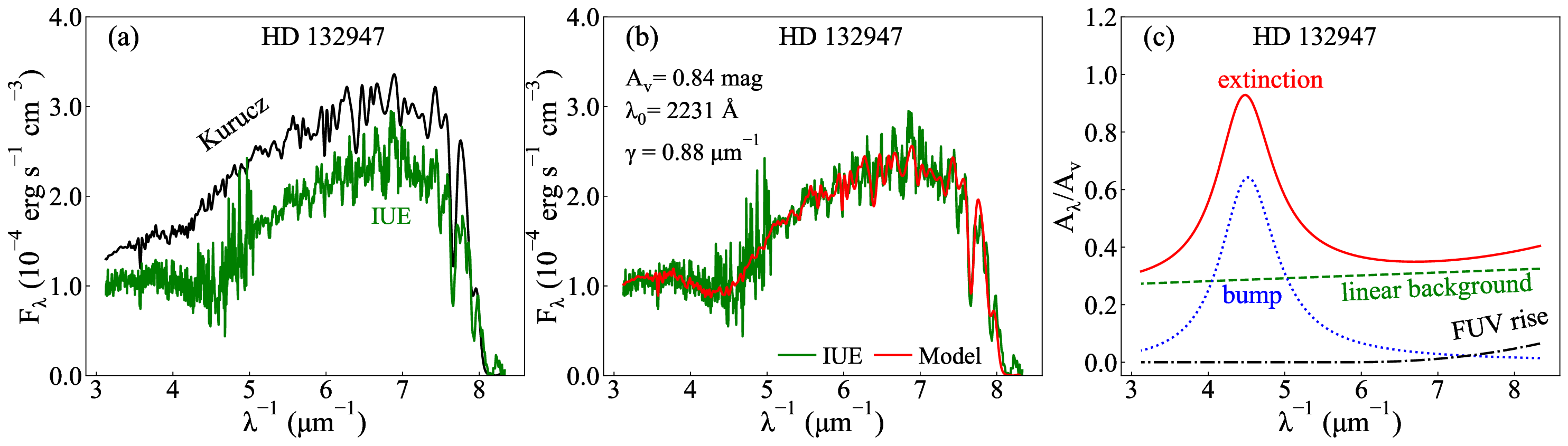}
\includegraphics[width=16cm,angle=0]{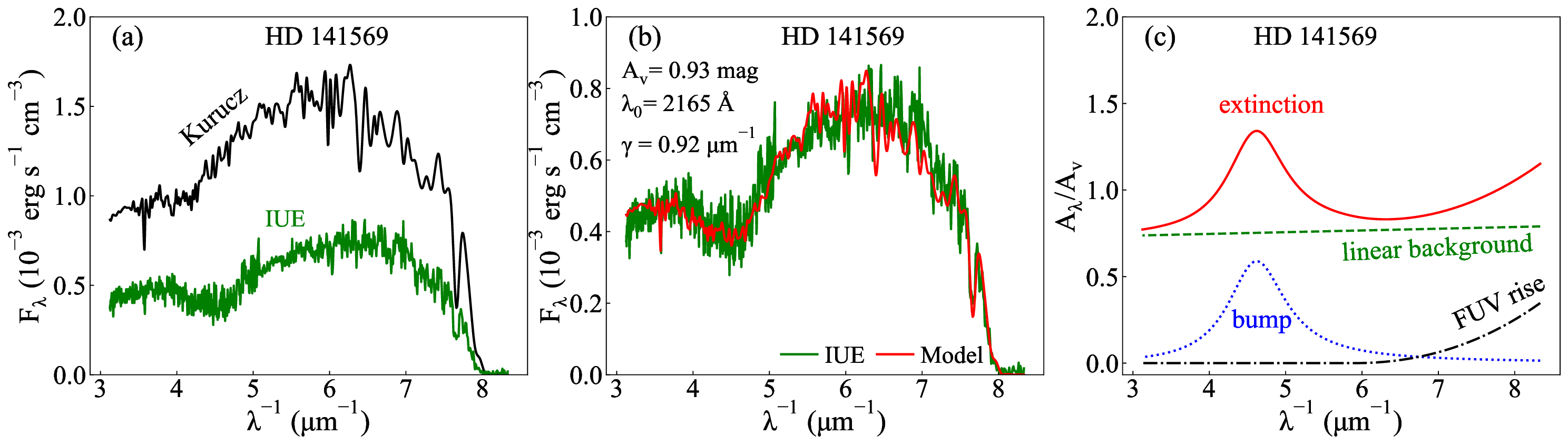}
\includegraphics[width=16cm,angle=0]{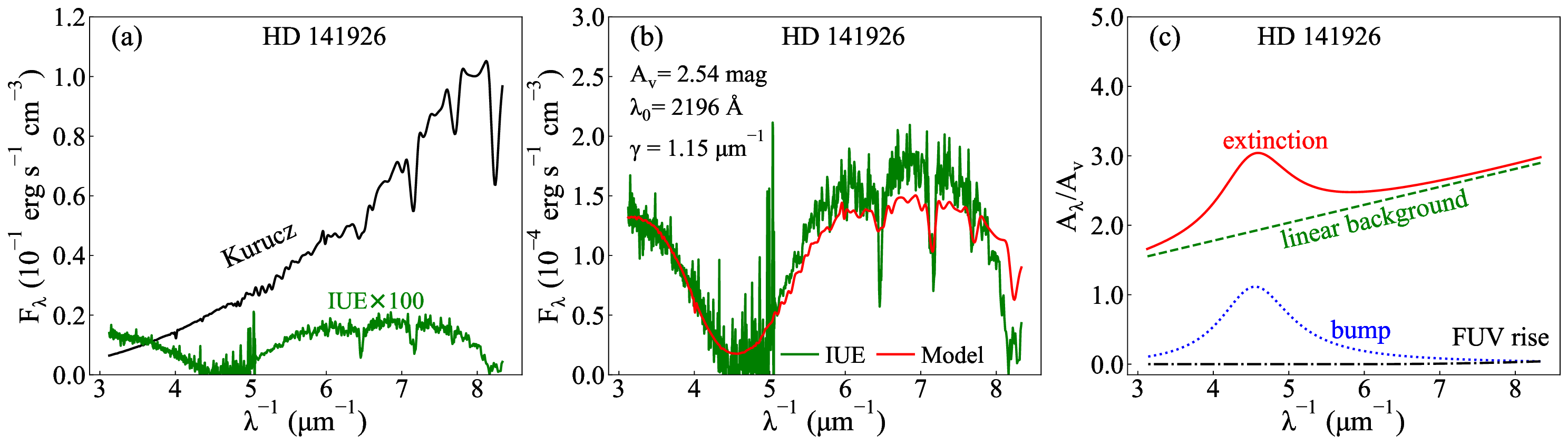}
\includegraphics[width=16cm,angle=0]{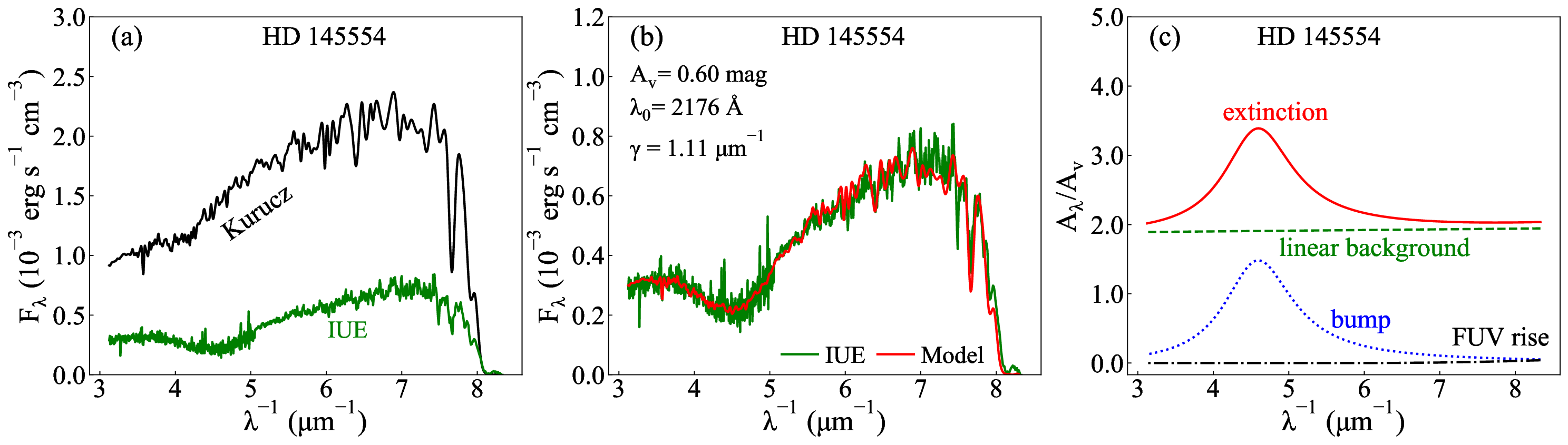}
\end{center}
\vspace{-5mm}
\caption{\label{fig:modfit3} 
  Same as Figure~\ref{fig:modfit1}
  but for HD~100546, HD~132947, HD~141569, HD~141926, and HD~145554.
	 }
\vspace{-3mm}
\end{figure*}

\begin{figure*}
\vspace{-3mm}
\begin{center}
\includegraphics[width=16cm,angle=0]{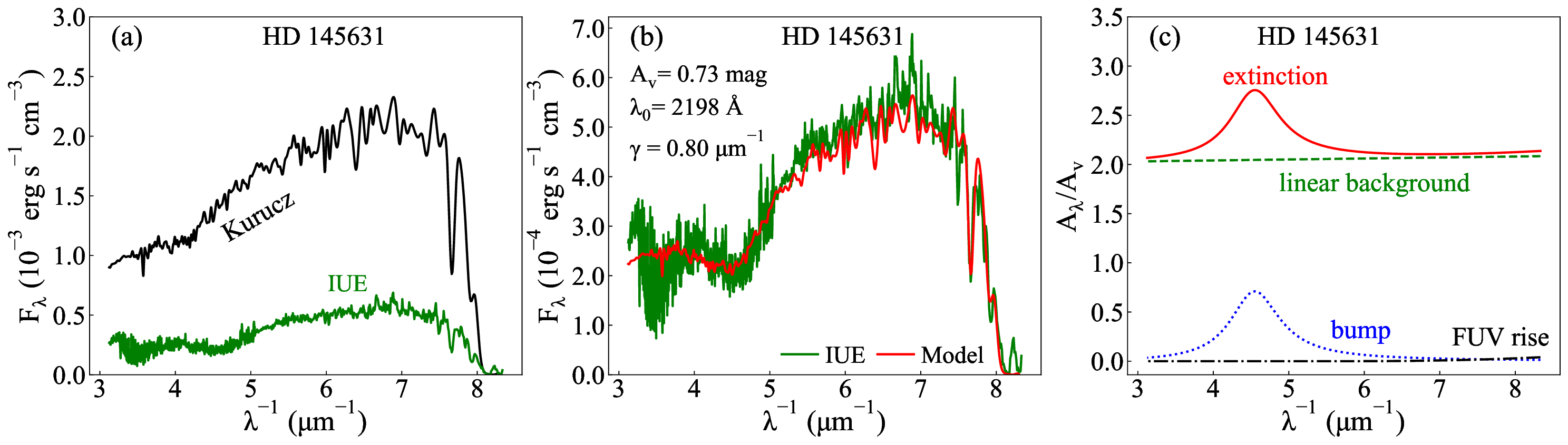}
\includegraphics[width=16cm,angle=0]{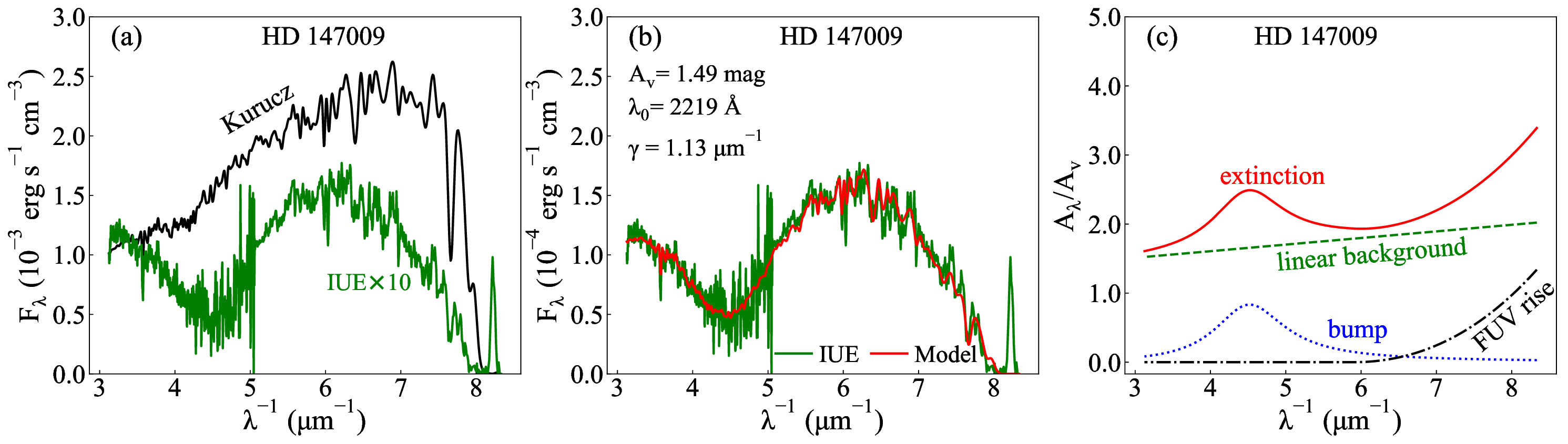}
\includegraphics[width=16cm,angle=0]{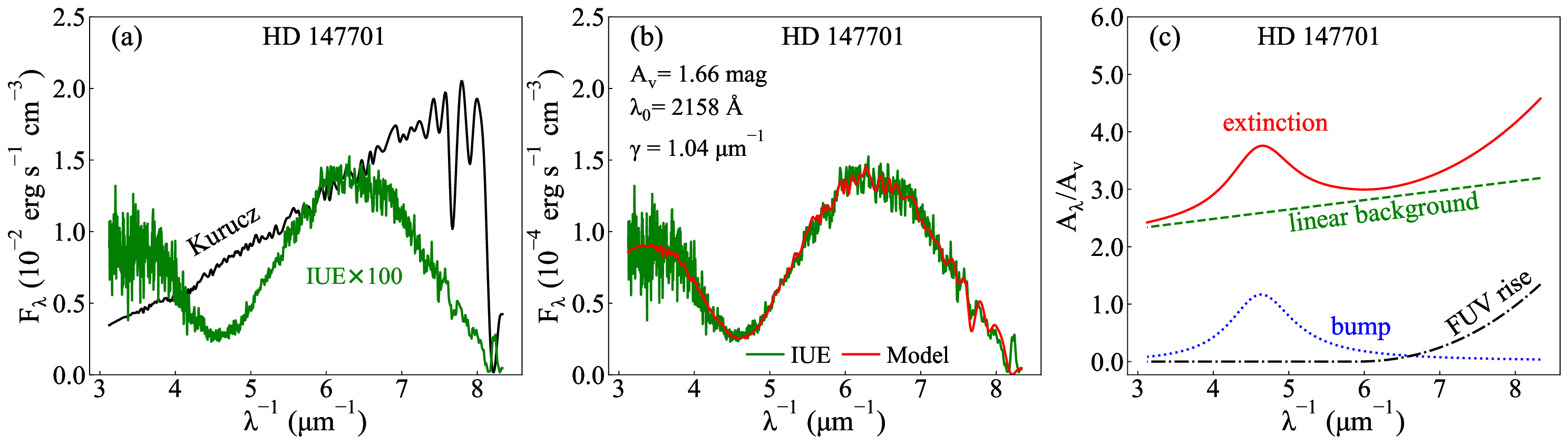}
\includegraphics[width=16cm,angle=0]{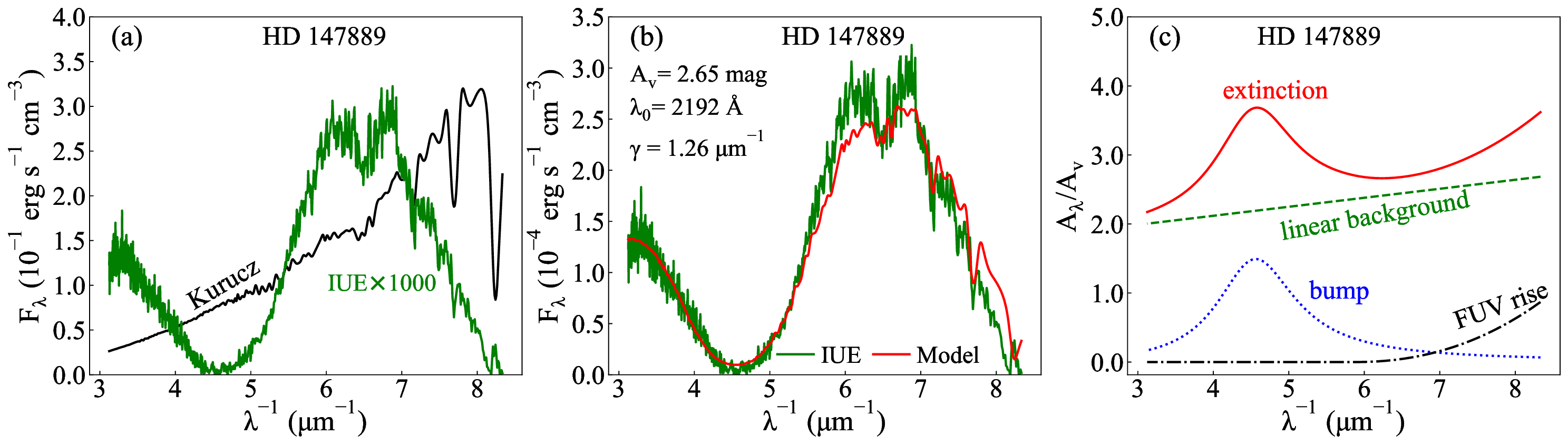}
\includegraphics[width=16cm,angle=0]{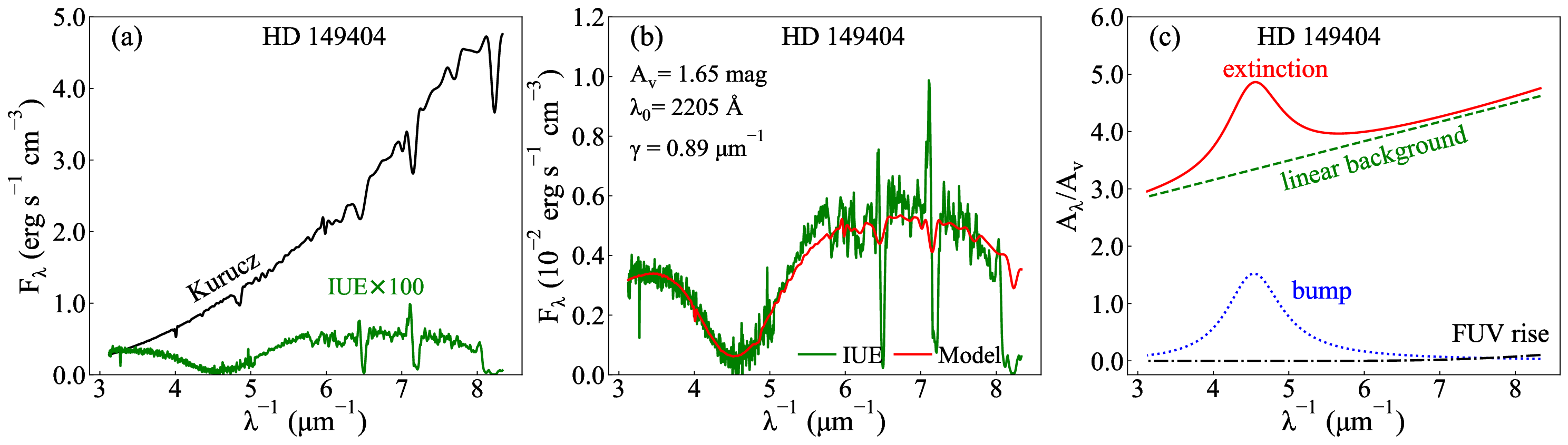}
\end{center}
\vspace{-5mm}
\caption{\label{fig:modfit4} 
  Same as Figure~\ref{fig:modfit1}
  but for HD~145631, HD~147009, HD~147701, HD~147889, and HD~149404.
	 }
\vspace{-3mm}
\end{figure*}

\begin{figure*}
\vspace{-3mm}
\begin{center}
\includegraphics[width=16cm,angle=0]{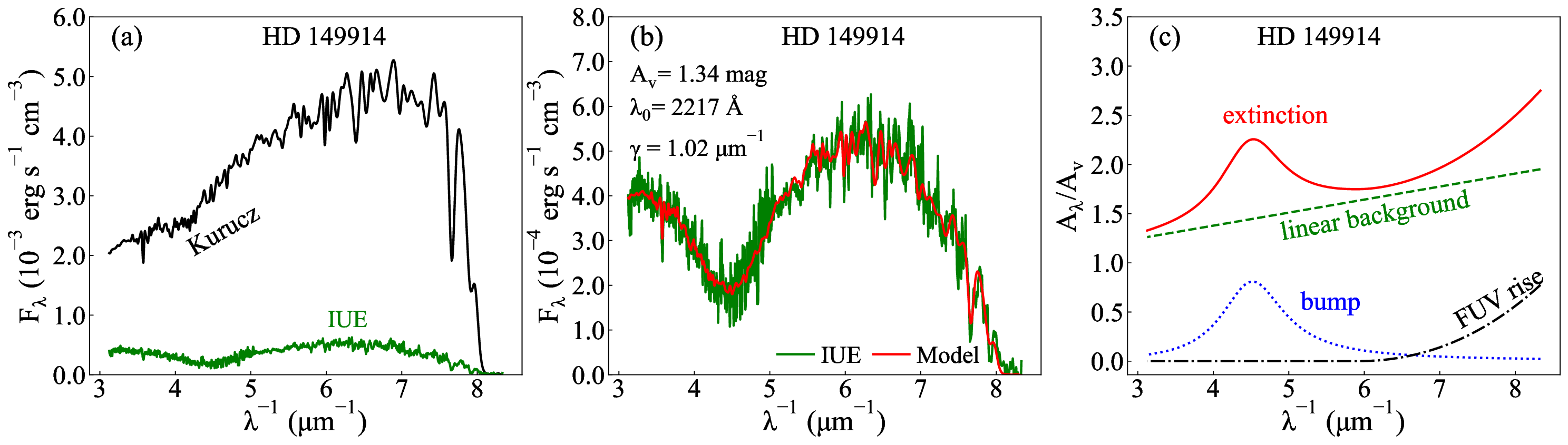}
\includegraphics[width=16cm,angle=0]{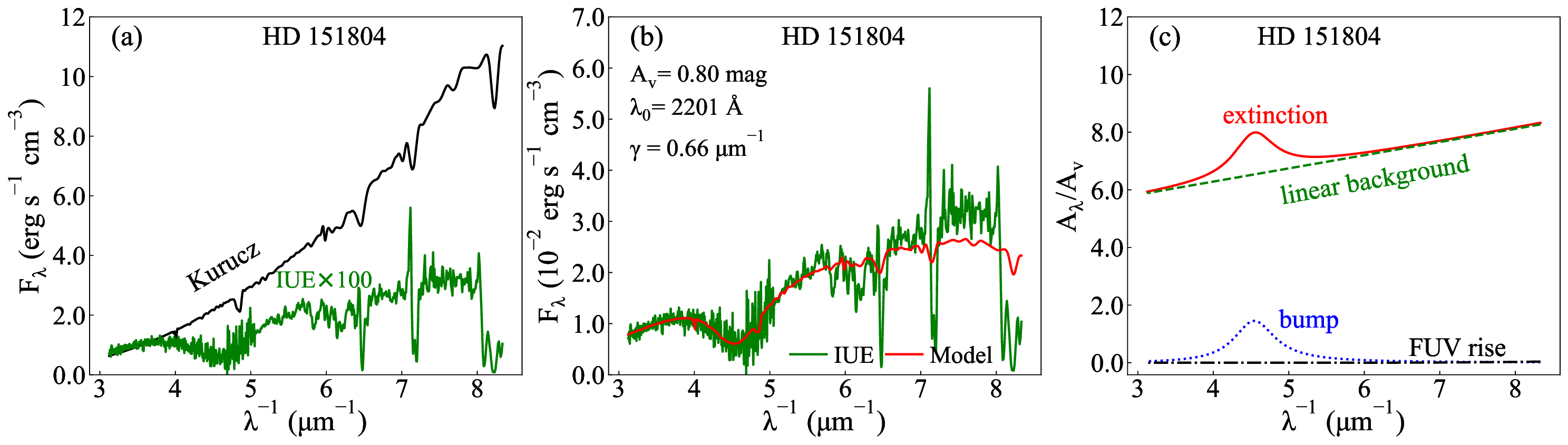}
\includegraphics[width=16cm,angle=0]{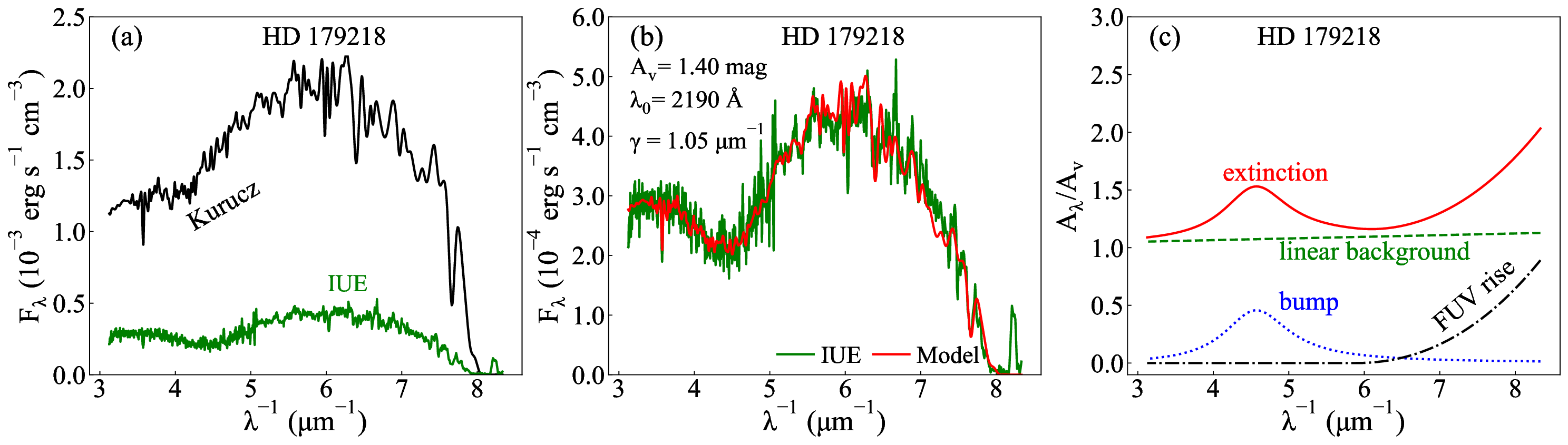}
\includegraphics[width=16cm,angle=0]{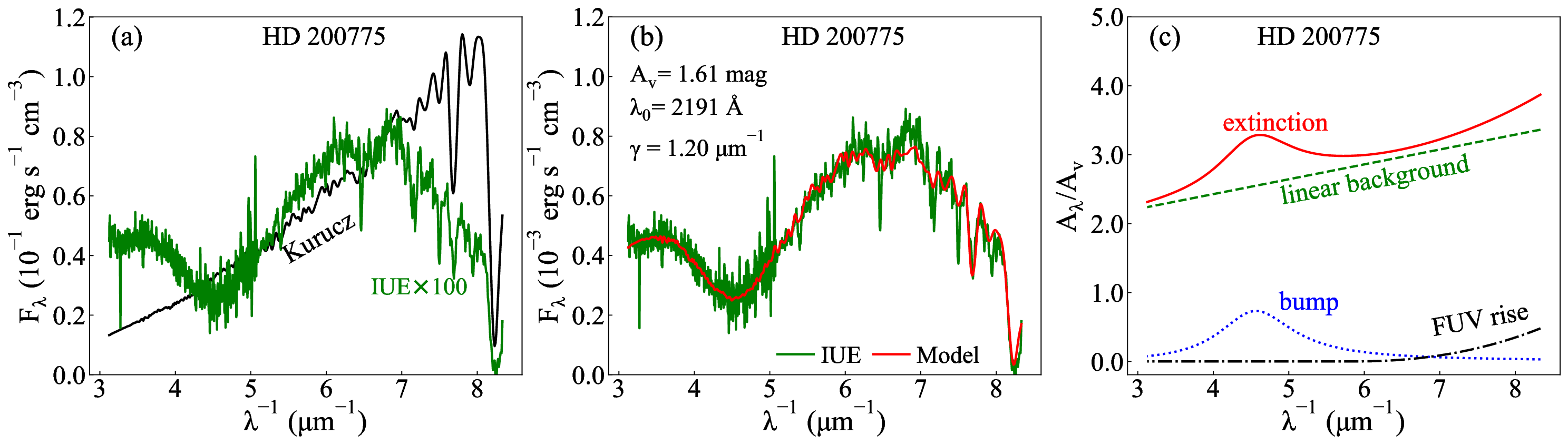}
\includegraphics[width=16cm,angle=0]{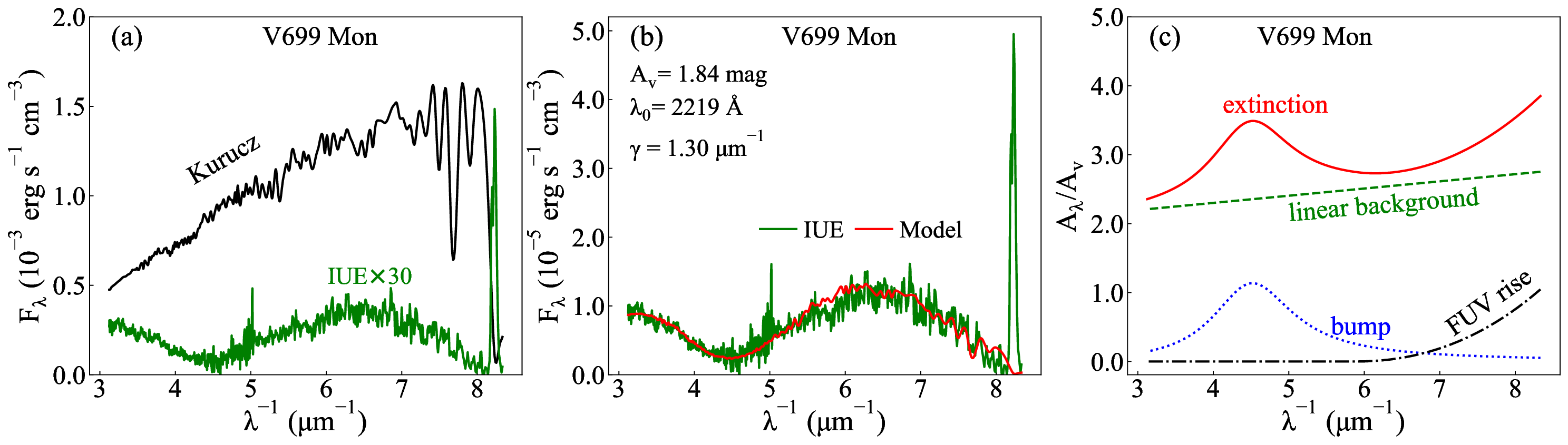}
\end{center}
\vspace{-5mm}
\caption{\label{fig:modfit5}
  Same as Figure~\ref{fig:modfit1}
  but for HD~149914, HD~151804, HD~179218, HD~200775, and V699~Mon.
	 }
\vspace{-3mm}
\end{figure*}

\begin{figure*}
\vspace{-3mm}
\begin{center}
\includegraphics[width=16cm,angle=0]{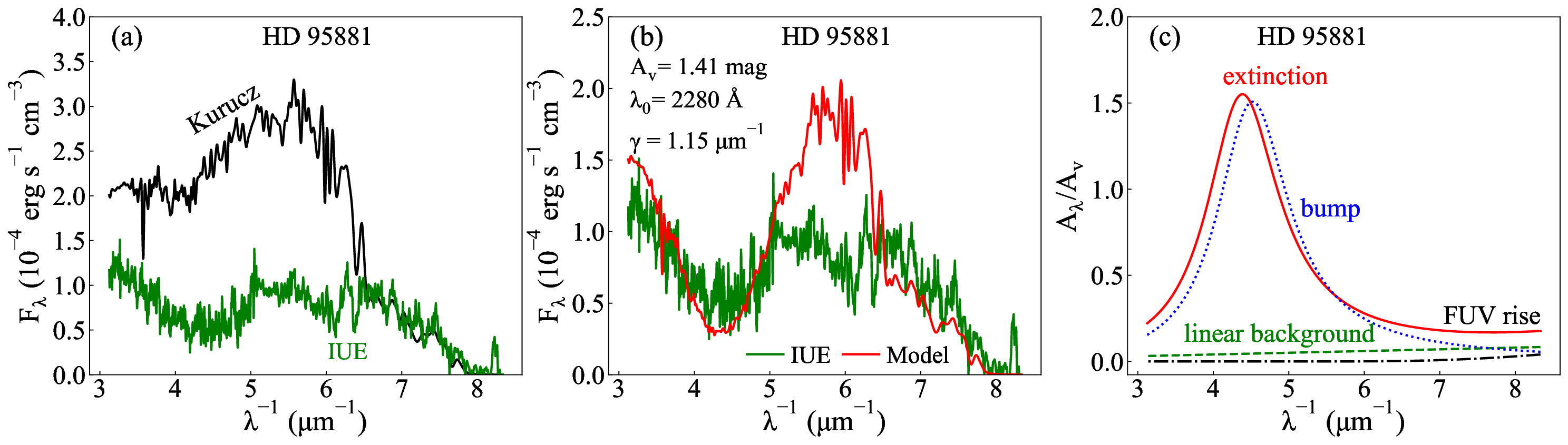}
\includegraphics[width=16cm,angle=0]{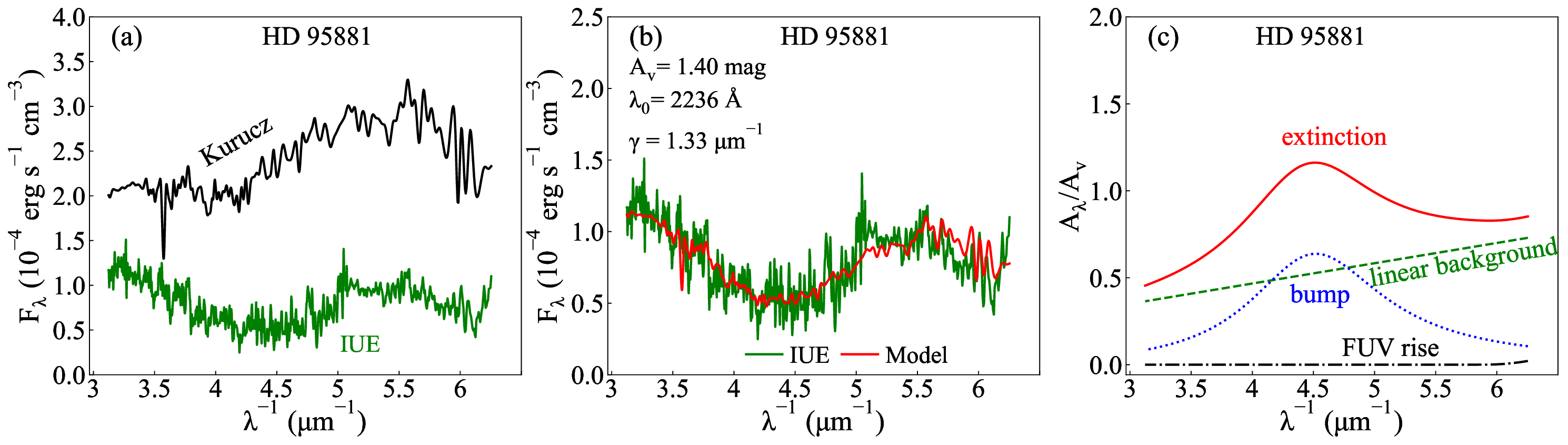}

\end{center}
\vspace{-5mm}
\caption{\label{fig:modfit6} 
  Same as Figure~\ref{fig:modfit1}
  but for the observed stellar and ``intrinsic'' spectra
  as well as the UV extinction curve of HD~95881
  in the 1200--3200$\Angstrom$ wavelength range
  (upper panel) and in the 1600--3200$\Angstrom$
  wavelength range (bottom panel).
  	 }
\end{figure*}

For each object, we show in
Figures~\ref{fig:modfit1}--\ref{fig:modfit6}
the observed IUE spectrum ($\Fobs$)
in comparison with the Kurucz model-based,
extinction-free ``intrinsic'' spectrum ($\Fintrinsic$),
and with the best-fit, dust-obscured
Kurucz model spectrum
($\Fintrinsic\exp\left\{-A_\lambda/1.086\right\}$).
Also shown are, for each line of sight,
the derived extinction curve
and its three components, i.e.,
the linear background,
the 2175$\Angstrom$ bump, and the FUV rise.

Figures~\ref{fig:modfit1}--\ref{fig:modfit6}
demonstrate that our approach is successful
in closely reproducing the obseved IUE spectra
in the wavelength range of
$\simali$1200--3200$\Angstrom$
for 25 (of 26) sources.
The only exception is HD~95881.
As shown in Figure~\ref{fig:modfit6},
our approach could not fit the observed IUE
spectrum in the wavelength range of
$\simali$1200--3200$\Angstrom$.
Our model spectrum at
$\simali$1600--2000$\Angstrom$
appreciably exceeds the observed IUE spectrum.
Nevertheless, if we confine ourselves to
1600--3200$\Angstrom$,
our model has no problem in
closely explaining the observed spectrum.
We have also tried to fit the IUE spectra
of all other 25 sources {\it exclusively}
in the 1600--3200$\Angstrom$ wavelength range.
It is found that the derived peak wavelength
and width of the extinction bump are
essentially the same as that derived
from fitting the IUE spectra
over the entire 1200--3200$\Angstrom$
wavelength range (see Table~\ref{tab:modpara}).

\begin{figure*}
\begin{center}
\includegraphics[width=10cm,angle=0]{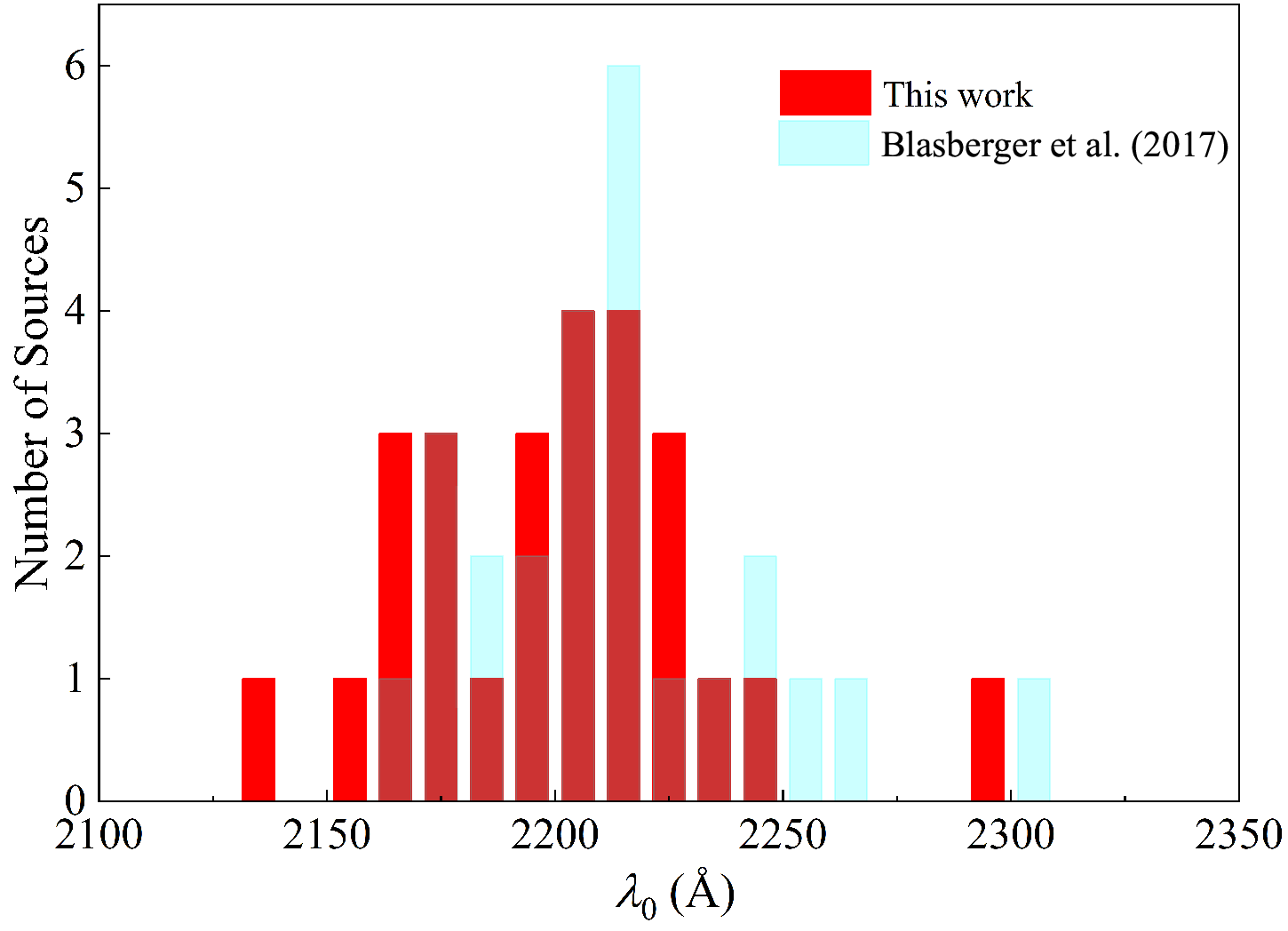}
\end{center}
\vspace{-5mm}
\caption{\label{fig:histogram} 
    Histograms of the central wavelengths of
    the 2175$\Angstrom$ extinction bump
    for the 26 lines of sight determined
    in this work (red bars) as well as that derived
    by Blasberger et al.\ (2017; cyan bars).
    }
\vspace{-3mm}
\end{figure*}

We show in Figure~\ref{fig:histogram}
the histogram of the central wavelengths
($\lambda_0$) of the 2175$\Angstrom$
extinction bump derived for all 26 sources.
Except for HD~95881 for which $\lambda_0$
is derived from fitting the IUE spectrum
over $\simali$1600--2000$\Angstrom$,
for all other 25 sources, we determine $\lambda_0$
from the IUE spectra over 1200--3200$\Angstrom$.
Figure~\ref{fig:histogram} reveals that
the peak wavelengths of the extinction bump
of our sample of 26 sources cluster around
2200$\Angstrom$, with a standard deviation
of only $\simali$33$\Angstrom$.
The line of sight toward HD~100546,
with $\lambda_0\approx 2299\Angstrom$,
has the longest peak wavelength.
If we exclude HD~100546,
the median peak wavelength
of our sample becomes
$\lambda_0\approx 2201\pm27\Angstrom$.
Figure~\ref{fig:histogram} demonstrates that
although the bumps of our sample peak
at relatively longer wavelengths,
their peak wavelengths are rather stable
and the scatters are small.
Compared with that derived
by Blasberger et al.\ (2017),
the peak wavelengths derived here
are appreciably less scattered.

\begin{figure*}
 \vspace{-3mm}
  \begin{center}
\includegraphics[width=12cm,angle=0]{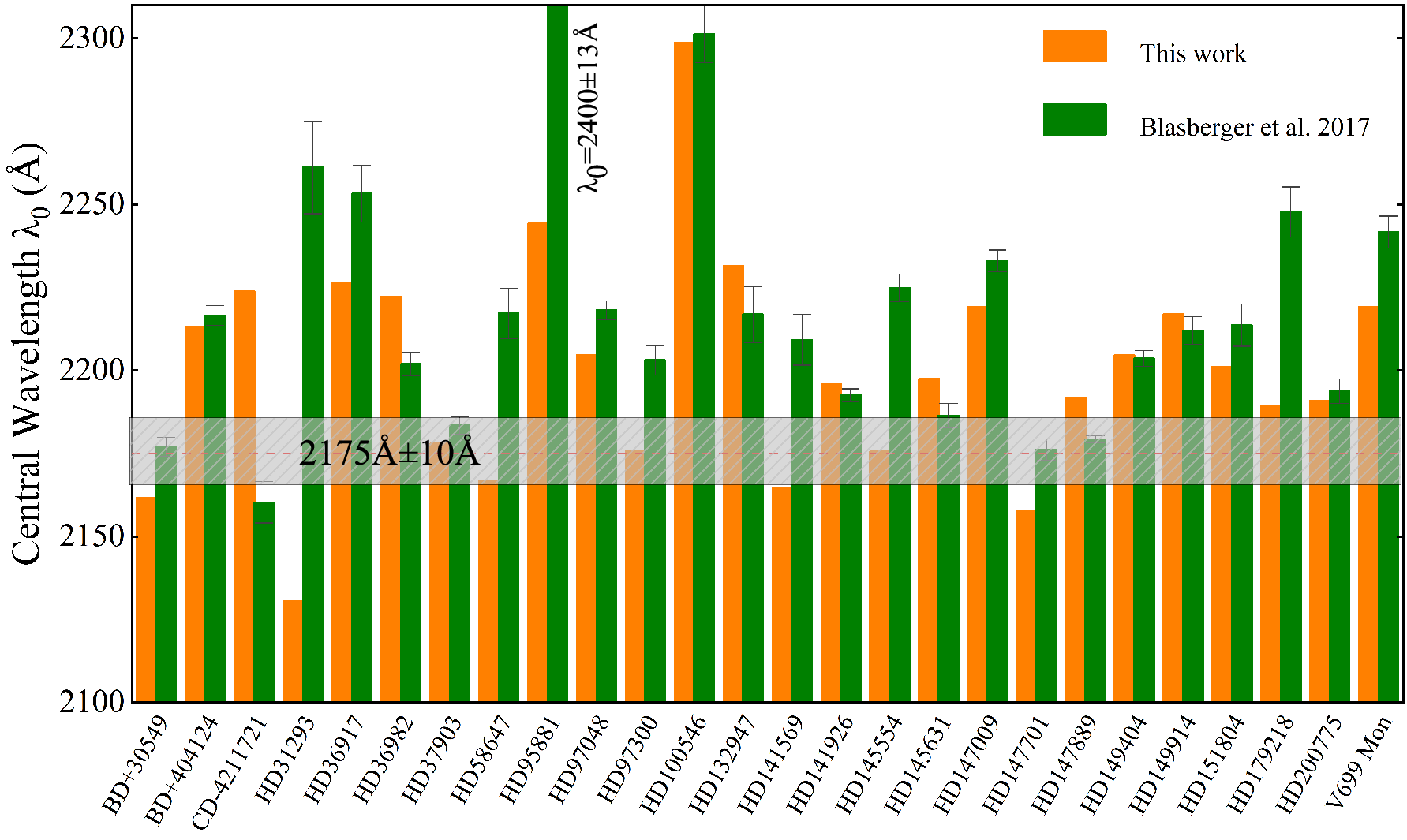}
  \end{center}
\vspace{-5mm}
\caption{\label{fig:comparison} 
  Comparison of the central wavelengths
  of the 2175$\Angstrom$ bump
  derived in this work (orange bars)
  with that determined
  by Blasberger et al.\ (2017; green bars).
  The horizontal bar shows the nominal
  wavelength of the 2175$\Angstrom$
  extinction bump and its
  nominal variation range.
	 }
\vspace{-3mm}
\end{figure*}

To determine the peak wavelengths of
the extinction bump of these sources,
Blasberger et al.\ (2017) took an approach
differing from ours. They fitted the IUE spectra
around the 2175$\Angstrom$ extinction bump
in the wavelength range of 1600--3200$\Angstrom$.
They took the Castelli \& Kurucz (2004) model
atmospheric spectrum ($\Frk$),
modulated by a power law to represent
the broadband extinction, and absorbed
by a feature with a Drude profile.
They assumed the measured flux spectrum
$\Fobs$ as a function of wavelength $\lambda$
to take the following form:
\begin{equation}
\label{eq:blasberger}
\Fobs = \Frk\times N \times
\left( \frac{\lambda}{2000 {\rm \AA } } \right)^{-\alpha}
\times \exp \left( \frac{-A}{\pi}\frac{\lambda^2}
{(\lambda^2-\lambda_0^2) +\lambda^2w^2} \right) ~,
\end{equation}
where $N$ is the normalization factor,
$\alpha$ is the power law slope, and
in the exponent, $A$ represents the amplitude
of the absorption feature, $w$ its width,
and $\lambda_0$ its central wavelength.
Comparing the approach of
Blasberger et al.\ (2017)
with ours (see eq.\,\ref{eq:fit}),
it is apparent that the major difference
is that they assumed the attenuation
caused by the broadband extinction
(i.e., the linear background extinction
component $c_1+c_2 \lambda^{-1}$)
take a power-law form $\lambda^{-\alpha}$,
while we take a physical approach
in which the attenuation caused
by the linear background extinction
is $\exp\left\{-c_1+c_2\lambda^{-1}\right\}$.
Apparently, it has to be justified
if the power-law $\lambda^{-\alpha}$
is a valid approximation of the attenuation
$\exp\left\{-c_1+c_2\lambda^{-1}\right\}$.
In Figure~\ref{fig:comparison} we compare
the peak wavelengths of the extinction bumps
derived here with that of Blasberger et al.\ (2017).
Clearly, the peak wavelengths derived
by Blasberger et al.\ (2017) were systematically
longer than that derived here
from a more physical approach.

\begin{figure*}
 \vspace{-3mm}
  \begin{center}
\includegraphics[width=12cm,angle=0]{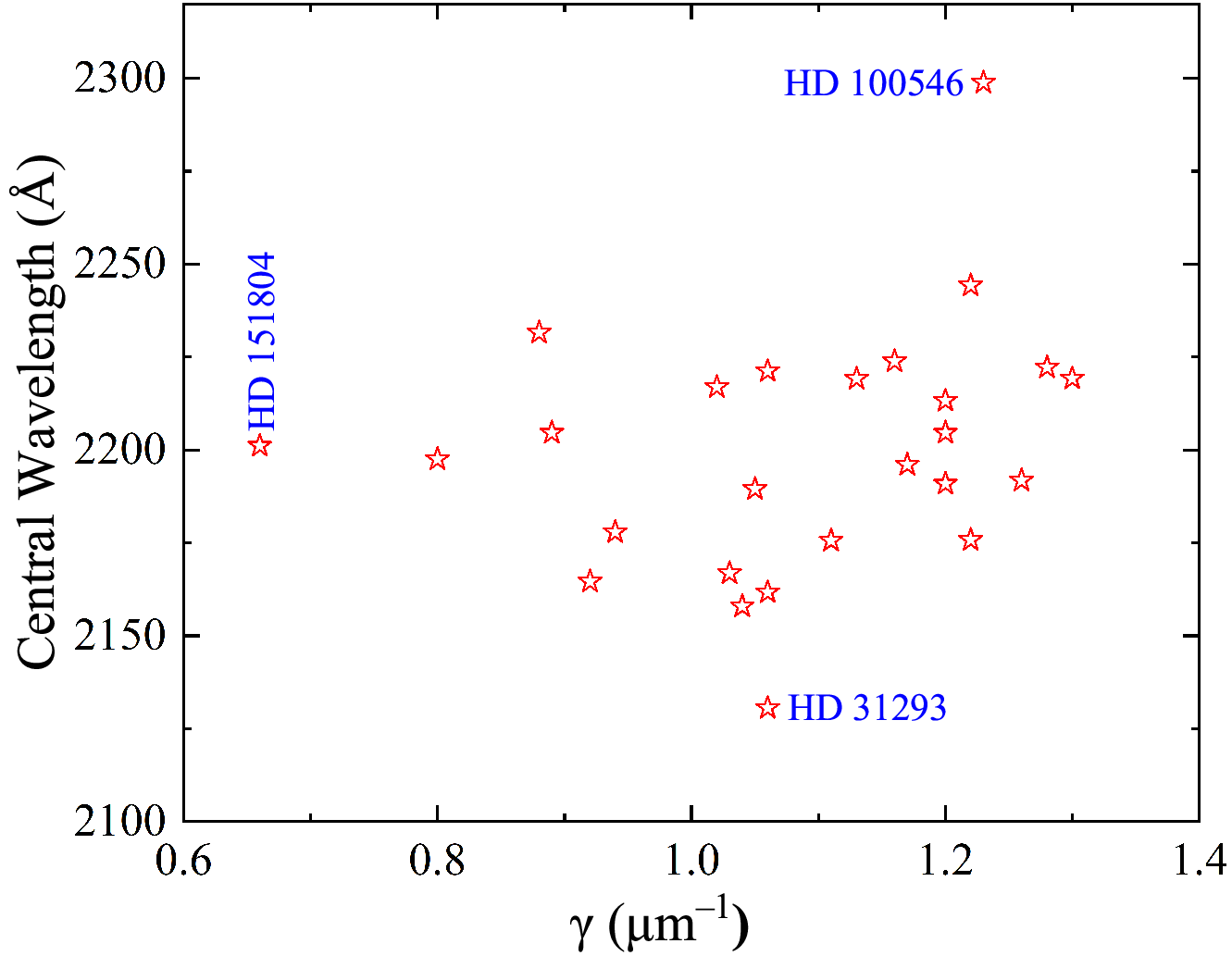}
  \end{center}
\vspace{-5mm}
\caption{\label{fig:wave_width} 
        The central wavelength of
        the 2175$\Angstrom$ extinction bump
        against its width.
	 }
\vspace{-3mm}
\end{figure*}

We argue that, based on a physical fitting
of the observed IUE spectra of 26 interstellar
lines of sight, the peak wavelengths of
the extinction bump are more or less stable
around $\simali$2200$\Angstrom$,
although they are longer than the nominal
2175$\Angstrom$. We note that, very recently,
an extinction bump peaking at
$\simali$2263$^{+20}_{-24}\Angstrom$
was discovered by JWST
in JADES-GS+53.15138-27.81917,
a galaxy at $z\approx6.71$ (Witstok et al.\ 2023).
This is notably longer than the nominal
2175$\Angstrom$.
Draine \& Malhotra (1993) have investigated
the effects on the 2175$\Angstrom$ extinction
bump of changes in graphite size and shape.
They found that, while small graphite grains
do produce a pronouced extinction bump,
the central wavelength of the bump is correlated
with the bump width.
In Figure~\ref{fig:wave_width}
we compare the central wavelengths
with the widths of the extinction bumps
derived here. Contrary to Draine \& Malhotra (1993),
no correlation is found. This is in agreement with
Fitzpatrick \& Massa (1986) who found large
variations in width of the extinction bump
with minimal, and uncorrelated, variations
in the central wavelength.

Finally, we note that, although the extinction
bump positions derived here are considerably
less scattered than that of Blasberger et al.\ (2017),
they are still appreciably more scattered than
that of the classical studies.
Fitzpatrick \& Massa (1986) studied the IUE data
of 45 stars and derived a stable bump peak
of $\simali$2174.4$\Angstrom$ and
a $2\sigma$ scatter of $\simali$10$\Angstrom$.
Based on the IUE data of 417 stars,
Valencic et al.\ (2004) placed the peak
of the extinction bump at $\simali$2179$\Angstrom$
with a $1\sigma$ scatter of only $\simali$5$\Angstrom$.
The discrepency between the present work
and that by Fitzpatrick \& Massa (1986)
and Valencic et al.\ (2004) may arise from
the validity of HAeBe stars as an interstellar
extinction indicator.
%
%
It is well recognized that HAeBe stars are surrounded
by circumstellar disks. Free-free emission from these
disks extends well into the V-band, thus adding to
the measured $E(B-V)$ value normally attributed to
interstellar reddening (e.g., see Garrison 1978,
Schild 1978, Witt \& Cottrell 1980).

Furthermore, HAeBe stars are frequently embedded
in nebulosity. When they are used for extinction studies,
scattered nebular continuum could enter
the relatively large apertures
($\simali$10$^{\prime\prime}$$\times$20$^{\prime\prime}$)
of the IUE spectrometers, adding to the apparent stellar
flux and changing the stellar spectral energy distribution (SED);
also, a substantial fraction of the dust-related extinction
may arise in the nebulosity itself, where the dust properties
may not be representative of typical interstellar dust.
It is not entirely clear why the nebulosity
and the local extinction in the vicinity of
HAeBe stars cause the bump positions to
systematically shift to longer wavelengths.
We argue that this is probably related to
the non-standard scattering and extinction
properties of the local (i.e., nebular and
circumstellar) dust. Due to grain growth,
the nebular and circumstellar dust grains
around HAeBe stars are generally much larger
than interstellar grains (e.g., see Li \& Lunine 2003)
and therefore they cause less obscuration in the UV.
This would effectively raise the flux level observed
at shorter wavelengths and therefore cause
the bump to shift to longer wavelengths.

Moreover, depending on the geometries and orientations
of the circumstellar disks surrounding those HAeBe stars,
the observed SEDs of the stars may suffer additional changes
due to absorption in the disk. In particular, unresolved
(at IUE resolution) absorptions in Fe\,II/Fe\,III multiplets
arising in the disk can alter the apparent SEDs of HAeBe stars
in the $\simali$2000--3500$\Angstrom$ wavelength region,
potentially affecting the appearance of
the 2175$\Angstrom$ extinction bump.

Finally, many HAeBe stars are actively accreting material
from the circumstellar disk. Accretion shocks from the infalling
material impacting on the stellar photosphere can give rise to
UV excess radiation that is more readily observed in cooler
accreting stars but which can also alter the intrinsic
photospheric SEDs of hotter stars.

In view of these caveats of HAeBe stars as
an effective indicator for interstellar extinction,
the discovery of variable peak positions
by Blasberger et al.\ (2017)
for the 2175$\Angstrom$ extinction bump
is questionable.
We should stress that, even if HAeBe stars
were well-suited for interstellar extinction studies,
as demonstrated earlier in this paper,
our physical approach leads to a much less
scattered extinction bump,
also indicating that the discovery of
variable peak positions
by Blasberger et al.\ (2017)
is questionable.

\section{Summary}\label{sec:summary}
We have determined the UV extinction curves
for a sample of interstellar lines of sight
toward 26 Herbig Ae/Be stars, by comparing
the observed IUE spectrum with the Kurucz
stellar model atmospheric spectrum
for each object. The extinction is represented
by an analytical formula consisting of
a linear background, a Drude function
for the extinction bump,
and a FUV nonlinear rise.
It is found that, peaking at $\simali$2200$\Angstrom$,
the peak positions of the extinction bumps
are rather stable and the scatters are small.
This is consistent with the conventional wisdom
that while the strength and width
of the interstellar extinction bump
vary with environment, its central wavelength
is quite invariant.

\section*{Acknowledgements}
We thank B.T.~Draine, Q.~Li, Q.~Lin
and the anonymous referee
for valuable suggestions.
QW and XJY are supported in part by
NSFC 12122302 and 11873041.
AL is supported in part by NASA grants
80NSSC19K0572 and 80NSSC19K0701.

\section*{Data Availability}
The data underlying this article will be shared
on reasonable request to the corresponding authors.


%

\begin{table*}
\caption[]{\footnotesize
                Stellar Parameters
                }
\label{tab:stellarpara}
\centerline{\footnotesize
\begin{tabular}{lcccccccc}
\noalign{\smallskip} \hline \hline \noalign{\smallskip}
  Object  &  Spectral Type$^\dagger$  &  E(B-V)$ ^\dagger$ &  $\Teff$  &
                                                     $L_{\star}/L_{\odot}$
  &  Distance  & Stellar Radius &  $R_{\star}/R_{\odot}$  &
                                                            References$^\ddagger$\\
   &    &  (mag)  &  (K)  & & (pc)  & (cm)  &  & \\

\noalign{\smallskip} \hline \noalign{\smallskip}
BD+30549  &  B8:p  &  0.59  &  12022  &  12.59  &  350  &  5.67E+10  &   &  (6)\\
BD+404124  &  B2Ve  &  0.98  &  22000  &  5900  &  980  &  3.67E+11  &   &  (1)\\
CD-4211721  &  B0IVe  &  1.58  &  30000  &  8710  &  400  &  2.39E+11  &   &  (5)\\
HD 31293  &  A0Ve  &  0.13  &  9800  &  57.5  &  144  &  1.83E+11  &  2.62  &  (3)\\
HD 36917  &  B9III/IV  &  0.17  &  10000  &  245.5  &  375  &  3.62E+11  &   &  (1)\\
HD 36982  &  B1.5Vp  &  0.37  &  20000  &  1659.59  &  375  &  2.38E+11  &  3.42  &  (3)\\
HD 37903  &  B3IV  &  0.31  &  23677  &  3311.31  &  820  &  2.37E+11  &   &  (7)\\
HD 58647  &  B9IV  &  0.14  &  10500  &  912  &  546  &  6.33E+11  &   &  (1)\\
HD 95881  &  A1/A2III/IV  &  0.13  &  9000  &  7.6  &  118  &  7.91E+10  &   &  (1)\\
HD 97048  &  A0Vep  &  0.31  &  10000  &  44  &  150  &  1.53E+11  &   &  (1)\\
HD 97300  &  B9V  &  0.24  &  10700  &  37  &  188  &  1.23E+11  &   &  (1)\\
HD 100546  &  B9Vne  &  0.48  &  10500  &  32  &  103  &  1.19E+11  &   &  (1)\\
HD 132947  &  B9V  &  0.13  &  10500  &  93.3  &  565  &  2.02E+11  &  3.1  &  (4)\\
HD 141569  &  B9.5V  &  0.09  &  9800  &  30.9  &  116  &  1.34E+11  &   &  (1)\\
HD 141926  &  B2nne  &  0.74  &  28000  &  20118.7  &  1254  &  6.75E+11  &  9.7  &  (2)\\
HD 145554  &  B9V  &  0.20  &  10471  &  42.66  &  137  &  1.38E+11  &   &  (7)\\
HD 145631  &  B9V  &  0.20  &  10471  &  44.67  &  141  &  1.41E+11  &   &  (7)\\
HD 147009  &  A0V  &  0.28  &  10471  &  42.66  &  130  &  1.38E+11  &   &  (7)\\
HD 147701  &  B5III  &  0.71  &  15488  &  204.17  &  140  &  1.38E+11  &   &  (7)\\
HD 147889  &  B2III/IV  &  1.05  &  21877  &  1995.26  &  140  &  2.16E+11  &   &  (6)\\
HD 149404  &  O8.5Iab(f)p  &  0.69  &  32000  &   &  810  &  2.36E+12  &  33.9  &  (8)\\
HD 149914  &  B9.5IV  &  0.26  &  10471  &  128.82  &  159  &  2.39E+11  &   &  (7)\\
HD 151804  &  O8.5Iab(f)p  &  0.36  &  34000  &   &  580  &  2.36E+12  &  34  &  (8)\\
HD 179218  &  A0Ve  &  0.11  &  9640  &  182  &  254  &  3.35E+11  &   &  (1)\\
HD 200775  &  B2Ve  &  0.61  &  18600  &  8912.5  &  429  &  6.30E+11  &   &  (1)\\
V699 Mon  &  B7IIne  &  0.70  &  13800  &  616.6  &  800  &  2.99E+11  &  4.3  &  (2)\\
\hline
\noalign{\smallskip}
\noalign{\smallskip} \noalign{\smallskip}
\end{tabular}
}
\begin{description}
\item[$^\dagger$] Taken from Blasberger et al.\ (2017).
\item[$^\ddagger$] Seok \& Li (2017), (2) Verhoeff et al. (2012),
                               (3) Alecian et al. (2013),
                               (4) Fairlamb et al.\ (2015),
                               (5) Boersma et al.\ (2009),
                               (6) Hamaguchi et al.\ (2005),
                               (7) Hern\'andez et al.\ (2005),
                               (8) Lamers et al.\ (1995).
%
\end{description}
\end{table*}

\begin{table*}
\caption[]{
    Extinction Model Parameters
    }
\label{tab:modpara}
\centerline{\footnotesize
\begin{tabular}{lcccccccc}
\noalign{\smallskip} \hline \hline \noalign{\smallskip}
Object  &  $A_V$  &  $x_0$  &  $\gamma$  &  $c_1$
            & $c_2$ & $c_3$  &  $c_4$  & $\mathcal{X}^2$\\
& (mag)  &  ($\Angstrom$)  &  ($\mu$m$^{-1}$)  &
              &   &  &  & \\
\noalign{\smallskip} \hline \noalign{\smallskip}
BD+30549  &  1.32 ± 0.00  &  2161.7 ± 0.01  &  1.06 ± 0.01  &  0.14 ± 0.02  &  0.23 ± 0.00  &  1.15 ± 0.01  &  0.39 ± 0.03  &  6.39\\
BD+404124  &  2.68 ± 0.00  &  2213.2 ± 0.01  &  1.20 ± 0.02  &  1.24 ± 0.03  &  0.20 ± 0.01  &  1.95 ± 0.01  &  0.15 ± 0.11  &  157.25\\
CD-4211721  &  2.34 ± 0.00  &  2223.8 ± 0.01  &  1.16 ± 0.04  &  2.00 ± 0.14  &  0.50 ± 0.04  &  1.99 ± 0.02  &  0.01 ± 1.46  &  128.07\\
HD 31293  &  0.92 ± 0.00  &  2130.6 ± 0.03  &  1.06 ± 0.03  &  -0.15 ± 0.03  &  0.34 ± 0.01  &  0.50 ± 0.02  &  0.10 ± 0.04  &  13.87\\
HD 36917  &  0.50 ± 0.00  &  2221.3 ± 0.02  &  1.06 ± 0.02  &  2.98 ± 0.07  &  0.02 ± 0.01  &  1.40 ± 0.05  &  0.01 ± 0.04  &  6.82\\
HD 36982  &  1.31 ± 0.00  &  2222.2 ± 0.01  &  1.28 ± 0.01  &  1.79 ± 0.01  &  0.01 ± 0.00  &  1.20 ± 0.01  &  0.09 ± 0.01  &  6.48\\
HD 37903  &  1.21 ± 0.00  &  2177.8 ± 0.00  &  0.94 ± 0.01  &  0.36 ± 0.01  &  0.12 ± 0.00  &  0.66 ± 0.00  &  0.10 ± 0.01  &  1.72\\
HD 58647  &  1.00 ± 0.00  &  2166.8 ± 0.01  &  1.03 ± 0.01  &  0.48 ± 0.02  &  0.05 ± 0.00  &  0.57 ± 0.01  &  0.01 ± 0.01  &  6.13\\
HD 95881  &  0.65 ± 0.00  &  2244.2 ± 0.01  &  1.22 ± 0.01  &  -0.18 ± 0.05  &  0.31 ± 0.01  &  2.00 ± 0.03  &  0.01 ± 1.81  &  10.67\\
HD 97048  &  1.43 ± 0.00  &  2204.6 ± 0.01  &  1.20 ± 0.01  &  1.25 ± 0.02  &  0.16 ± 0.00  &  0.93 ± 0.01  &  0.04 ± 0.02  &  6.58\\
HD 97300  & 1.53 ± 0.00  &  2204.6 ± 0.01  &  1.22 ± 0.01  &  1.34 ± 0.01  &  0.07 ± 0.00  &  0.93 ± 0.01  &  0.16 ± 0.01  &  7.07\\
HD 100546  &  0.99 ± 0.01  &  2298.8 ± 0.02  &  1.23 ± 0.02  &  0.14 ± 0.02  &  0.01 ± 0.00  &  0.68 ± 0.01  &  0.01 ± 0.01  &  4.82\\
HD 132947  &  0.84 ± 0.01  &  2231.5 ± 0.01  &  0.88 ± 0.01  &  0.24 ± 0.02  &  0.01 ± 0.00  &  0.50 ± 0.01  &  0.02 ± 0.01  &  5.26\\
HD 141569  &  0.93 ± 0.00  &  2164.6 ± 0.01  &  0.92 ± 0.01  &  0.71 ± 0.02  &  0.01 ± 0.00  &  0.50 ± 0.01  &  0.09 ± 0.01  &  5.1\\
HD 141926  &  2.54 ± 0.00  &  2201.9 ± 0.00  &  1.17 ± 0.01  &  0.68 ± 0.03  &  0.28 ± 0.01  &  1.46 ± 0.01  &  0.01 ± 0.08  &  175.57\\
HD 145554  &  0.60 ± 0.00  &  2175.6 ± 0.01  &  1.11 ± 0.01  &  1.86 ± 0.03  &  0.01 ± 0.00  &  1.81 ± 0.02  &  0.01 ± 0.01  &  4.62\\
HD 145631  &  0.73 ± 0.00  &  2197.5 ± 0.02  &  0.80 ± 0.02  &  2.00 ± 0.04  &  0.01 ± 0.01  &  0.45 ± 0.02  &  0.01 ± 0.02  &  28.17\\
HD 147009  &  1.49 ± 0.00  &  2219.1 ± 0.01  &  1.13 ± 0.01  &  1.23 ± 0.02  &  0.10 ± 0.00  &  1.07 ± 0.01  &  0.33 ± 0.02  &  5.47\\
HD 147701  &  1.66 ± 0.00  &  2157.9 ± 0.01  &  1.04 ± 0.01  &  1.83 ± 0.01  &  0.16 ± 0.00  &  1.25 ± 0.01  &  0.34 ± 0.03  &  7.41\\
HD 147889  &  2.65 ± 0.00  &  2191.8 ± 0.01  &  1.26 ± 0.04  &  1.60 ± 0.02  &  0.13 ± 0.00  &  2.39 ± 0.01  &  0.22 ± 0.04  &  38.89\\
HD 149404  &  1.65 ± 0.00  &  2204.6 ± 0.00  &  0.89 ± 0.01  &  1.81 ± 0.02  &  0.34 ± 0.01  &  1.21 ± 0.01  &  0.03 ± 0.04  &  10.07\\
HD 149914  &  1.34 ± 0.00  &  2216.9 ± 0.01  &  1.02 ± 0.01  &  0.85 ± 0.01  &  0.13 ± 0.00  &  0.84 ± 0.01  &  0.19 ± 0.02  &  6.44\\
HD 151804  &  0.80 ± 0.00  &  2201.1 ± 0.01  &  0.66 ± 0.01  &  4.46 ± 0.04  &  0.46 ± 0.01  &  0.64 ± 0.01  &  0.01 ± 0.03  &  11.89\\
HD 179218  &  1.40 ± 0.00  &  2189.5 ± 0.01  &  1.05 ± 0.01  &  1.01 ± 0.01  &  0.01 ± 0.00  &  0.50 ± 0.01  &  0.22 ± 0.01  &  3.47\\
HD 200775  &  1.61 ± 0.00  &  2190.9 ± 0.01  &  1.20 ± 0.01  &  1.57 ± 0.01  &  0.22 ± 0.00  &  1.05 ± 0.01  &  0.12 ± 0.02  &  5.48\\
V699 Mon  &  1.84 ± 0.00  &  2219.2 ± 0.01  &  1.30 ± 0.01  &  1.88 ± 0.02  &  0.10 ± 0.00  &  1.92 ± 0.01  &  0.26 ± 0.03  &  13.17\\
\hline
\noalign{\smallskip}
\noalign{\smallskip} \noalign{\smallskip}
\end{tabular}
}
\end{table*}

\bsp
\label{lastpage}

\begin{thebibliography}{30}
\expandafter\ifx\csname natexlab\endcsname\relax\def\natexlab#1{#1}\fi
\bibitem[]{}Alecian, E., Wade, G. A., Catala, C., et al.\ 2013, MNRAS, 429, 1001
\bibitem[]{}Blasberger, A., Behar, E., Perets, H. B., Brosch, N., \& Tielens, A. G. G. M.\ 2017, ApJ, 836, 173
\bibitem[]{}Boersma, C., Peeters, E., Mart\'in-Hern\'andez, N. L., et al.\ 2009, A\&A, 502, 175
\bibitem[]{}Cardelli, J. A., Clayton, G. C., \& Mathis, J. S.\ 1989, ApJ, 345, 245
\bibitem[]{}Castelli, F., \& Kurucz, R. L.\ 2004, arXiv:0405087
\bibitem[]{}Chhowalla, M., Wang, H., Sano, N., Teo, K.~B.,
            Lee, S.~B., \& Amaratunga, G.~A.\ 2003,
            Phys. Rev. Lett., 90, 155504
\bibitem[]{}Decleir, M., Gordon, K.~D., Andrews, J.~E., et al.\ 2022, ApJ, 930, 15
\bibitem[]{}Draine, B.T.\ 1988, ApJ, 333, 848
\bibitem[]{}Draine, B.T. \& Malhotra, S.\ 1993, ApJ, 414, 632
\bibitem[]{}Fairlamb, J. R., Oudmaijer, R. D., Mendigut\'a, I., Ilee, J. D., \& van den Ancker, M. E.\ 2015, MNRAS, 453, 976
\bibitem[]{}Fitzpatrick, E. L., \& Massa, D.\ 1986, ApJ, 307, 286
\bibitem[]{}Fitzpatrick, E. L., \& Massa, D.\ 1988, ApJ, 328, 734
\bibitem[]{}Garrison, L.~M.\ 1978, ApJ, 224, 535
\bibitem[]{}Gordon, K.~D., Clayton, G.~C., Decleir, M., et al.\ 2023, arXiv:2304.01991
\bibitem[]{}Hamaguchi, K., Yamauchi S., \& Koyama K.\ 2005, ApJ, 618, 360
\bibitem[]{}Hern\'andez, J., Calvet, N., Hartmann, L., et al.\ 2005, AJ, 129, 856
\bibitem[]{}Iglesias-Groth, S., Ruiz, A., Bret\'{o}n, J.,
                  \& Gomez Llorente, J.~M.\ 2003,
                  J. Chem. Phys., 118, 7103
\bibitem[]{}Joblin, C., L\'eger, A., \& Martin, P.\ 1992, ApJ, 393, L79
\bibitem[]{}Lamers, H. J. G. L. M., Snow, T. P., \& Lindholm, D. M.\ 1995, 455, 269
\bibitem[]{}Li, A., \& Draine, B.~T.\ 2001, ApJ, 554, 778
\bibitem[]{}Li, A., \& Lunine, J.I.\ 2003, ApJ, 594, 987
\bibitem[]{}Li, A., Chen, J.H., Li, M.P., Shi, Q.J.,
                 \& Wang, Y.J.\ 2008, MNRAS, 390, L39
\bibitem[]{}Ma, X.Y., Zhu, Y.Y., Yan, Q.B., You, J.Y.,
                 \& Su, G.\ 2020, MNRAS, 497, 2190
\bibitem[]{}Massa, D., Gordon, K.~D., \& Fitzpatrick, E.~L.\ 2022, ApJ, 925, 19
\bibitem[]{}Mathis, J.S.\ 1994, ApJ, 422, 176
\bibitem[]{}Ruiz, A., Bret\'{o}n, J.,
            \& Gomez Llorente, J.~M.\ 2005,
            Phys. Rev. Lett., 94, 105501
\bibitem[]{}Schild, R.~E.\ 1978, ApJ, 37, 77
\bibitem[]{}Seok, J. Y. \& Li, A.\ 2017, ApJ, 835, 291
\bibitem[]{}Sheng X.L., Yan Q.B., Ye F., Zheng Q.R.,
                  \& Su G.\ 2011, Phys. Rev. Lett., 106, 155703
\bibitem[]{}Stecher, T. P.\ 1965, ApJ, 142, 1683
\bibitem[]{}Stecher, T. P. \& Donn, B.\ 1965, ApJ, 142, 1681
\bibitem[]{}Steglich, M., J\"ager, C., Huisken, F., et al.\ 2013, ApJS, 208, 26
\bibitem[]{}Valencic, L. A., Clayton, G. C., \& Gordon, K. D.\
	         2004, ApJ, 616, 912
\bibitem[]{}Verhoeff, A.P., Waters, L.B.F.M.,
             van den Ancker, M.E., et al.\ 2012, A\&A, 538, A101
\bibitem[]{}Whittet, D.C B.\ 2022,
                 Dust in the Galactic Environment (3rd Edition),
                Bristol: IOP Publishing
\bibitem[]{}Witstok, J., Shivaei, I., Smit, R., et al.\ 2023,
                 Nature, in press (arxiv:2302.05468)
\bibitem[]{}Witt, A.~N., \& Cottrell, M.~J.\ 1980, ApJ, 235, 899
%

\end{thebibliography}
\end{document}